\documentclass[prd,12pt,a4paper,nofootinbib,tightenlines,showkeys]{revtex4}
\usepackage[left=2.5cm,right=2.5cm,top=2.5cm,bottom=2.5cm]{geometry}

\usepackage[frenchb,english]{babel}
\usepackage[latin1]{inputenc}
\usepackage{enumerate}
\usepackage{amsmath}
\usepackage{amsthm}
\usepackage{amssymb}
\usepackage{graphicx}
\usepackage{floatflt}
\usepackage{fancybox}
\usepackage{appendix}
\usepackage{enumitem}
\usepackage{subfigure}
\usepackage{array}

\usepackage[normalem]{ulem}

\usepackage{hyperref}
\usepackage[b]{esvect}

\usepackage{color}

\renewcommand{\Re}{\textrm{Re}}
\renewcommand{\Im}{\textrm{Im}}

\newcommand{\om}{\omega}

\newcommand{\lam}{\lambda}

\newcommand{\gam}{\gamma}

\newcommand{\eff}{{\rm eff}}

\newcommand{\m}{{\rm min}}
\newcommand{\M}{{\rm max}}

\newcommand{\p}{\partial}

\newcommand{\be} {\begin{equation}}
\newcommand{\ee} {\end{equation}}
\newcommand{\bsub}{\begin{subequations}}
\newcommand{\esub}{\end{subequations}}
\newcommand{\bea}{\begin{eqnarray}}
\newcommand{\eea}{\end{eqnarray}}
\newcommand{\bi} {\begin{itemize}}
\newcommand{\ei} {\end{itemize}}
\newcommand{\ben} {\begin{enumerate}}
\newcommand{\een} {\end{enumerate}}
\newcommand{\bmat} {\begin{pmatrix}}
\newcommand{\emat} {\end{pmatrix}}
\newcommand{\bal} {\begin{aligned}}
\newcommand{\eal} {\end{aligned}}
\newcommand{\btab}{\begin{tabular}}
\newcommand{\etab}{\end{tabular}}

\newcommand{\eq}[1]{equation~\eqref{#1}}

\begin{document}
\selectlanguage{english}

\title{Anomalous transmission through periodic resistive sheets}

\author{Antonin Coutant}
\email{antonin.coutant@univ-lemans.fr}
\affiliation{Laboratoire d'Acoustique de l'Université du Mans, Unite Mixte de Recherche 6613, Centre National de la Recherche Scientifique, Avenue O. Messiaen, F-72085 LE MANS Cedex 9, France}

\author{Yves Aurégan}
\email{yves.auregan@univ-lemans.fr}
\affiliation{Laboratoire d'Acoustique de l'Université du Mans, Unite Mixte de Recherche 6613, Centre National de la Recherche Scientifique, Avenue O. Messiaen, F-72085 LE MANS Cedex 9, France}

\author{Vincent Pagneux}
\email{vincent.pagneux@univ-lemans.fr}
\affiliation{Laboratoire d'Acoustique de l'Université du Mans, Unite Mixte de Recherche 6613, Centre National de la Recherche Scientifique, Avenue O. Messiaen, F-72085 LE MANS Cedex 9, France}

\date{\today}

\begin{abstract}
This work investigates anomalous transmission effects in periodic dissipative media, which is identified as an acoustic analogue of the Borrmann effect. For this, the scattering of acoustic waves on a set of equidistant resistive sheets is considered. It is shown both theoretically and experimentally that at the Bragg frequency of the system, the transmission coefficient is significantly higher than at other frequencies. The optimal conditions are identified: one needs a large number of sheets, which induce a very narrow peak, and the resistive sheets must be very thin compared to the wavelength, which gives the highest maximal transmission. Using the transfer matrix formalism, it is shown that this effect occurs when the two eigenvalues of the transfer matrix coalesce, i.e. at an exceptional point. Exploiting this algebraic condition, it is possible to obtain similar anomalous transmission peaks in more general periodic media. In particular, the system can be tuned to show a peak at an arbitrary long wavelength. 
\end{abstract}

\keywords{Acoustic wave propagation, Absorption, Phononic Crystals.}

\pacs{43.20.Mv, 
43.20.Fn, 
}

\maketitle

\section{Introduction}

Periodic media, also known as phononic crystals are now a widely used tool to control the propagation properties of sound waves~\cite{Deymier}. The first effect of periodicity is to induce band gaps: frequency ranges where waves become evanescent inside the material. This bands and gaps structure of the dispersion relation has found a large range of uses, among which large decreases of transmission~\cite{vasseur01}, wave localization and wave guiding~\cite{Kafesaki00,Kane14}, or high quality resonances~\cite{Kalozoumis18}. 

On the contrary, the effect of absorption in periodic media has received much less attention. Several authors have investigated the changes of the dispersion relation and whether absorption tends to enlarge or reduce band gaps~\cite{Psarobas01,Lee10,Hussein13}. However, in periodic media combination of absorption and Bragg scattering can lead to peculiar transmission properties. A well-known effect of that type in X-ray crystallography, called the Borrmann effect, is the anomalous transmission of waves across a crystal slab at the Bragg frequency~\cite{Borrmann41,Batterman64}. More recently, a similar anomalous transmission was observed in photonic crystals~\cite{Vinogradov09,Novikov17,Novikov19}. In these works, a light wave is sent on a slab made of successive pairs of layers made of a high-absorbing and a low-absorbing material. When the Bragg condition is met (the normal wavelength is twice the spatial periodicity of the medium), an anomalously high transmission is observed. In acoustics, a similar anomalous transmission has been reported for layered lossy structures in~\cite{Cebrecos14}. 

In this work, we study theoretically and experimentally an acoustic realisation of the Borrmann effect, using plane resistive sheets placed equidistantly. We show that there is a peak in transmission at the Bragg frequency (distance between successive sheets is half a wavelength), with a width decreasing rapidly with the number of sheets. This contrasts with what is usually observed in phononic crystals, where the Bragg frequency is associated to a low transmission. Moreover, we show that the anomalous transmission is the highest when the resistive sheets of a given resistance are thin enough compared to the wavelength. 

We also provide a detailed analysis of this effect by analyzing the Bloch wave solutions inside the material using the transfer matrix formalism. In particular, we show that the anomalous transmission corresponds to an \emph{exceptional point} (EP) of the $2 \times 2$ transfer matrix (normal incidence or monomodal assumption), where its two eigenvalues coalesce. This novel understanding of the Borrmann effect allows us to generalize it by placing resistive sheets inside an otherwise lossless periodic medium. If the sheets are placed adequately, one can obtain a Borrmann anomalous transmission at either one edge of the first band gap or the other. In particular, this allows us to construct configurations displaying a transmission peak at wavelengths arbitrarily larger than the system size. 

The paper is organized as follows. In section II we provide a theoretical description of the Borrmann effect using resistive sheets as well as its main properties. In section III we present the experimental characterization of the anomalous transmission. In section IV we show how this anomalous transmission can be obtained at different frequencies using resistive sheets inside a lossless periodic medium.

\section{Theoretical description of layered dissipative media}
\begin{figure}[!ht]
\centering
\includegraphics[width=0.75\columnwidth]{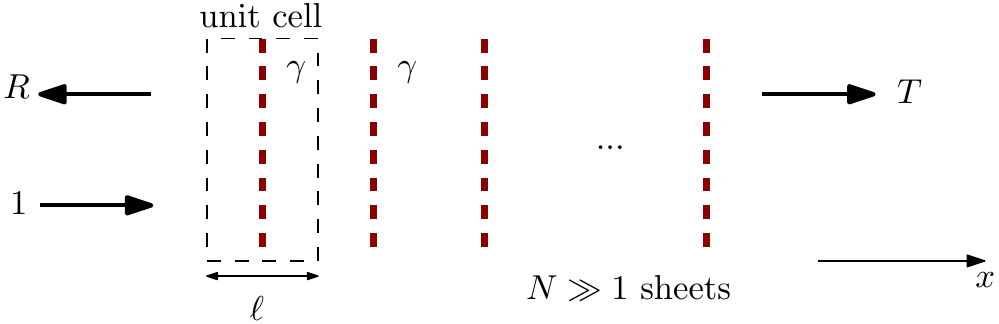}
\caption{Layered medium made of equidistantly spaced resistive sheets.} 
\label{WM_Schema_Fig} 
\end{figure}

We consider acoustic waves propagating in air. The pressure field obeys the Hemholtz equation 
\be \label{Heq}
\Delta p + k^2 p = 0, 
\ee
with $k$ the natural wavenumber. Time dependence $e^{-i \om t}$ is omitted, with $\om = k c_0$ and $c_0$ the speed of sound in air. The main ingredient of our setup is the use of purely resistive sheets~\cite{Ingard} made of a very fine-meshed fabric. These sheets have acoustical properties much like porous materials: the characteristic size of the holes are such that viscous effects dominate inside the fabric and the wave is attenuated~\cite{Allard}. Since their thickness is also much smaller than the wavelength, an acoustic wave across such a sheet undergoes a pressure jump from one side to another. The pressure discontinuity is characterized by the resistance $\gam$ such that 
\be \label{WM_jump}
[p]_{0^-}^{0^+} = - \rho_0 c_0 \gam u(0), 
\ee
where $u$ is the acoustic velocity, which is continuous across the sheet~\cite{Ingard}, and $\rho_0$ is the air density. Notice that the resistance $\gamma$ is adimensionalized by the natural impedance $\rho_0 c_0$ of air. We now investigate how acoustic waves propagate in a layered medium made of a number $N$ of identical resistive sheets equidistantly spaced by $\ell$ (see Fig.~\ref{WM_Schema_Fig}). 

\subsection{Periodic medium}
\label{WM_Bloch_Sec}
As a first step, we describe how waves propagate inside such a layered medium ($N\to \infty$). We consider for this a unit cell of size $\ell$, with a resistive sheet in the center (see Fig.~\ref{WM_Schema_Fig}). We assume the problem to be one-dimensional, either by sending plane waves with normal incidence, or because the sheets are placed in a waveguide of small cross section (monomodal propagation), as for the experiments of section~\ref{Exp_Sec}. In this case, it is very convenient to use the formalism of the transfer matrix~\cite{Soukoulis}. The entries of the transfer matrix are the forward and backward propagating components of the pressure field (see Appendix~\ref{Mmat_Smat_App} for details). The transfer matrix of a single cell is given by  
\bea \label{Mmat_def}
M_c &=& \bmat \vspace{8pt} t_1 - \dfrac{r_1^2}{t_1} & \dfrac{r_1}{t_1} \\ - \dfrac{r_1}{t_1} &  \dfrac{1}{t_1} \emat,  
\eea
where $t_1$ (resp. $r_1$) is the transmission (resp. reflection) coefficient of a single cell. Since such a cell (and hence also a collection of them) is reciprocal and mirror symmetric, the transmission and reflection are identical whether the incident wave comes from the left or the right~\cite{Soukoulis}. This justifies the specific form of the transfer matrix \eqref{Mmat_def}.  

\begin{figure}[!ht]
\centering
\includegraphics[width=0.95\columnwidth]{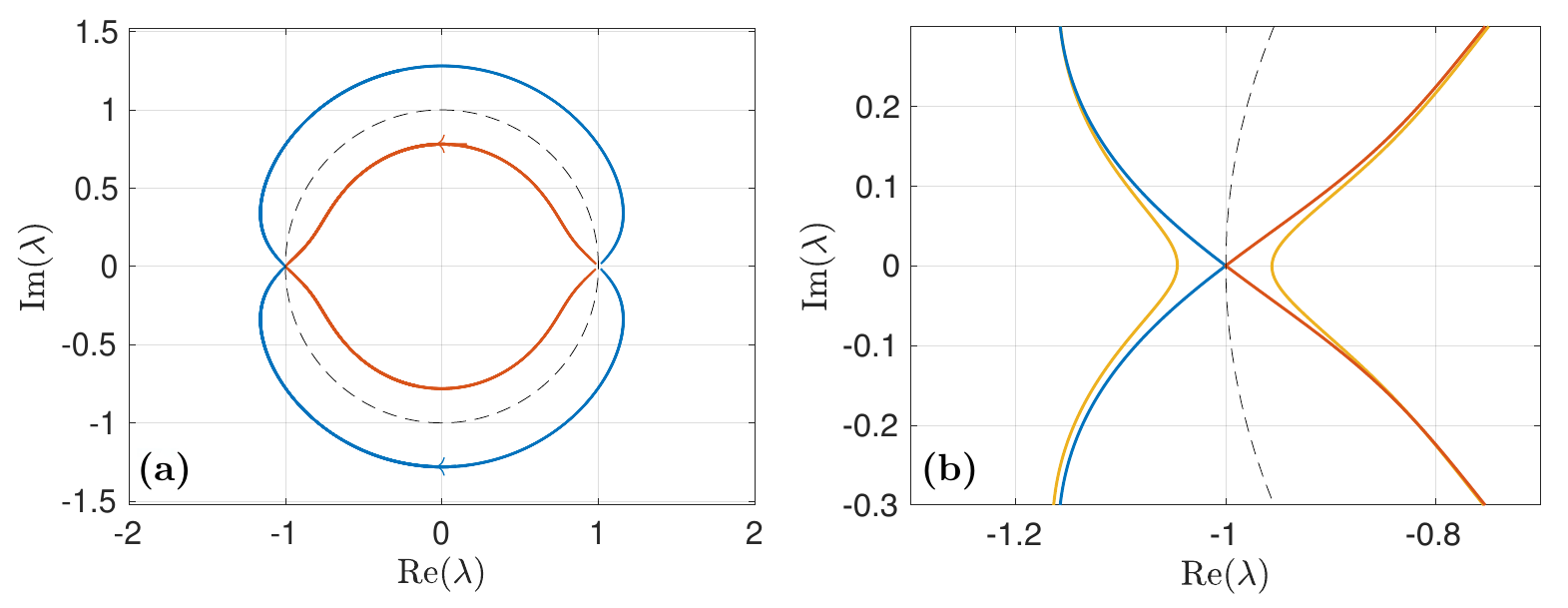}
\caption{(a) Eigenvalues $\lam = e^{\pm i q \ell}$ of the transfer matrix $M_c$ of a single cell in the complex plane. $k\ell$ runs from $0$ to $2\pi$, and we took $\gam=0.5$. The arrows show the direction of increasing wavenumber $k$ and the black dashed line is the unit circle. (b) Same as (a) but zoomed near the EP. We added the trajectory of  the eigenvalues for a sheet of finite thickness $\ell_0 = 0.1\ell$ (in yellow); this will be discussed in Sec.~\ref{Thick_Borr_Sec}.}
\label{WM_Bloch_Fig} 
\end{figure}

By solving the Hemholtz equation \eqref{Heq} from $-\ell/2$ to $\ell/2$ and using the jump condition \eqref{WM_jump}, the transfer matrix is given in terms of the resistance $\gam$, size $\ell$ and wavenumber $k$: 
\be \label{Mmat_cell}
M_c = \bmat \vspace{8pt} \left(1 - \dfrac{\gam}{2}\right) e^{i k \ell} & \dfrac{\gam}{2} \\ - \dfrac{\gam}{2} &  \left(1 + \dfrac{\gam}{2}\right) e^{-i k \ell} \emat. 
\ee
Eigenvectors of the transfer matrix of a cell represent Bloch wave solutions of the periodic system. The corresponding eigenvalue is the amplitude change after each cell. Since the system is reciprocal, the two eigenvalues are inverse, hence one can write them as $e^{\pm i q \ell}$, where $q$ is in general complex. This amounts to looking for solutions in the form of Bloch waves with effective wavenumber $q$: 
\be
p = e^{iqx} \varphi(x), 
\ee
where $\varphi$ has the periodicity of the medium $\varphi(x+\ell) = \varphi(x)$. Using the fact that the sum of both eigenvalues is the trace of $M_c$, we obtain the dispersion relation of the periodic medium: 
\be \label{Borr_DispRel}
2\cos(q\ell) = 2\cos(k\ell) - i \gam \sin(k\ell). 
\ee
To represent this relation, we show in Fig.~\ref{WM_Bloch_Fig} the trajectories of the eigenvalues of $M_c$ in the complex plane as the frequency $k$ varies. From \eq{Borr_DispRel}, we immediately notice a remarkable property of the constructed medium: at the Bragg frequencies $k\ell = \pi$ the last term in \eqref{Borr_DispRel} vanishes. In the following we focus on the Bragg frequency but we point out the same is true for all multiple of the Bragg frequency $k\ell = n \pi$ ($n \in \mathbb N$), and therefore so is the anomalous transmission we describe in this work. Equation~\eqref{Borr_DispRel} leads to two distinct behaviors for the absorption of waves. At generic frequencies $k\ell \neq n \pi$, $q$ has an imaginary part governed by $\gam$, and hence, a wave traveling in the medium in attenuated exponentially with the distance. On the contrary, at the Bragg frequencies, the imaginary part of $q$ vanishes, and one has a double solution for $q$ (see Fig.~\ref{WM_Bloch_Fig}). 

\begin{figure}[!ht]
\centering
\includegraphics[width=0.49\columnwidth]{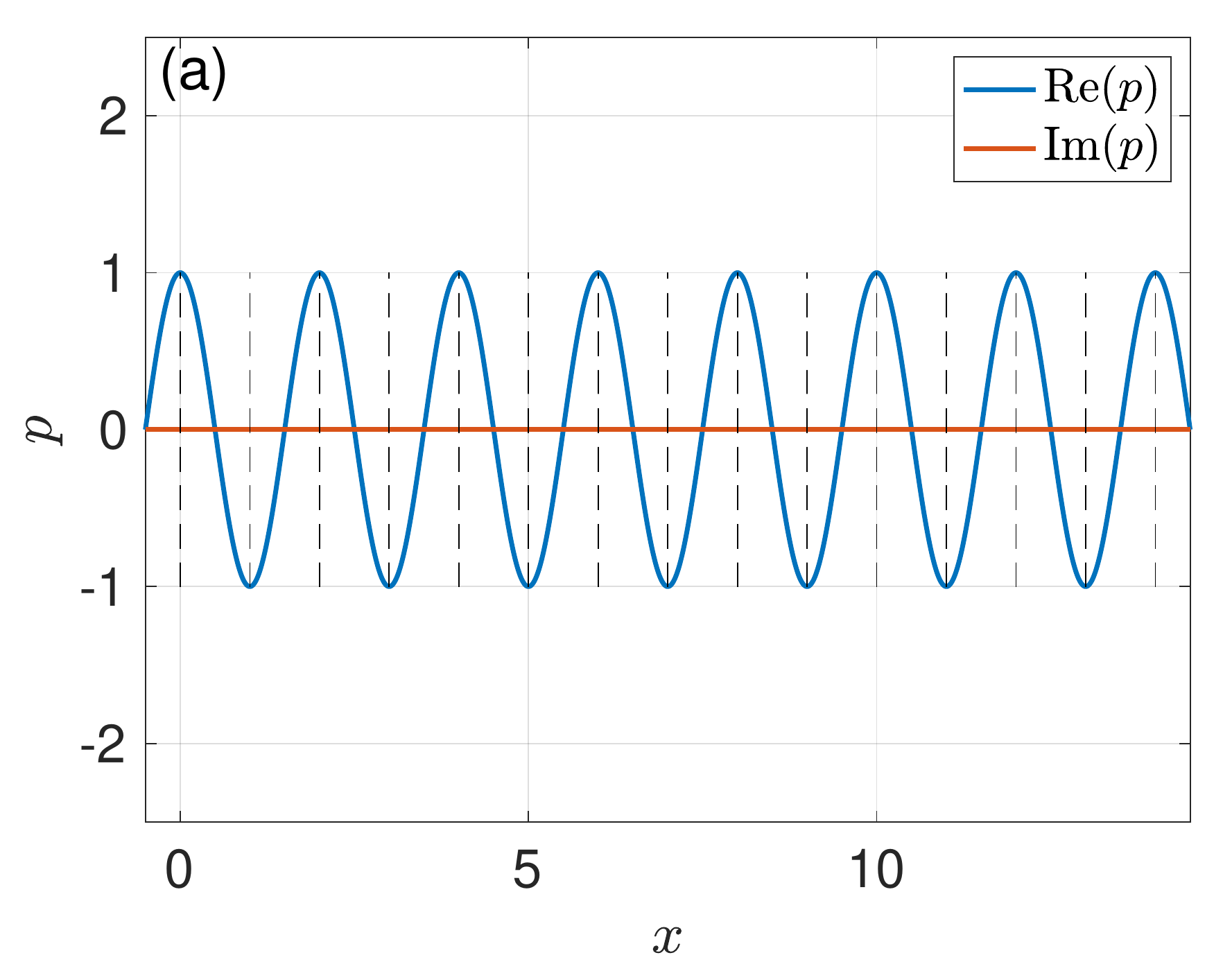}
\includegraphics[width=0.49\columnwidth]{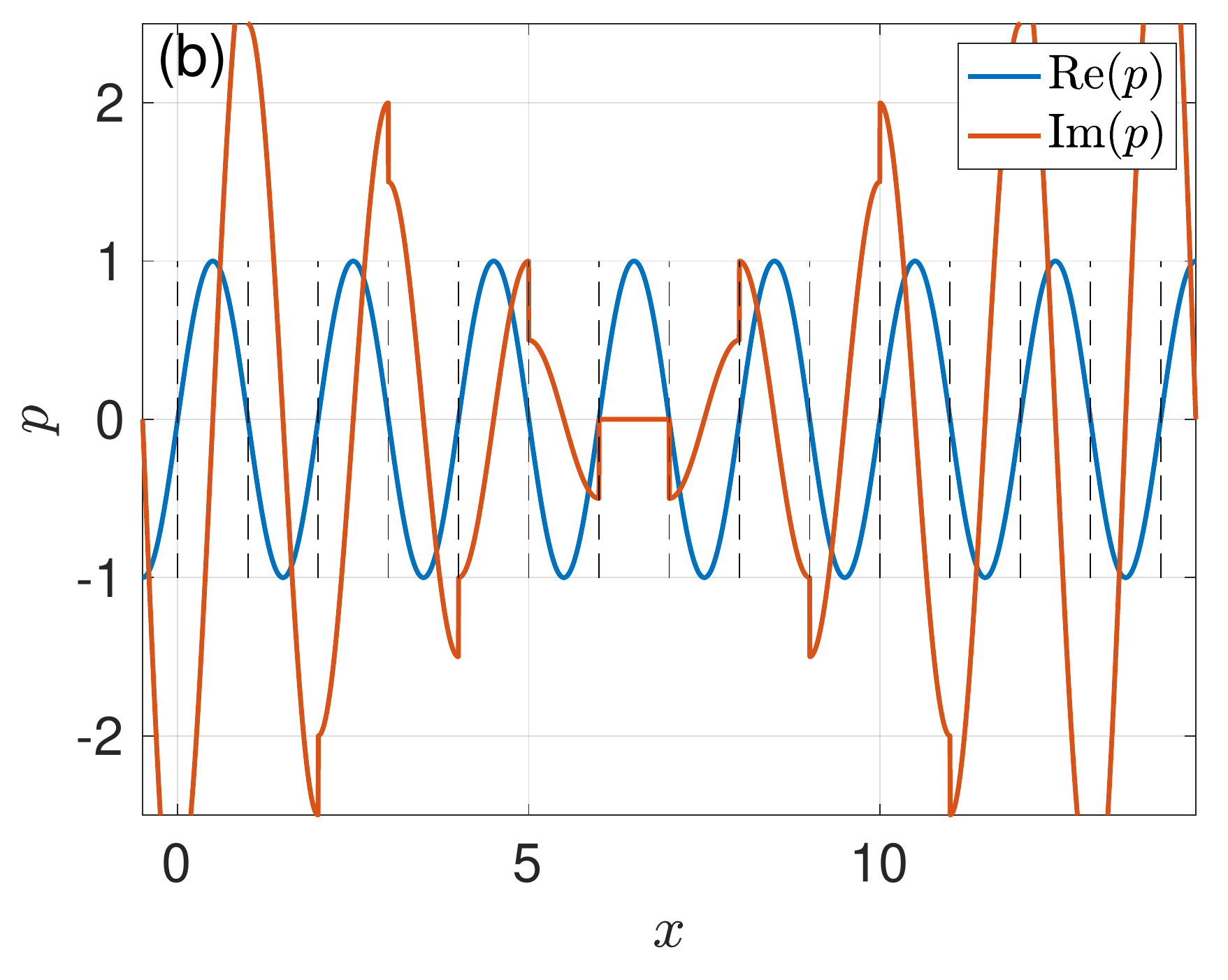}
\caption{Wave solutions at the Bragg frequency ($k\ell = \pi$) with $\gam=0.5$. The locations of the resistive sheets are shown as black dashed lines. (a) Eigen-vector of $M_c$. (b) Generalized eigen-vector of $M_c$.}
\label{WM_Bloch_Vec_Fig} 
\end{figure}

To fully understand the absorption properties at the Bragg frequency, we discuss in details the two linearly independent wave solutions. At the Bragg frequency, the $2\times 2$ transfer matrix $M_c$ possesses a single eigenvalue, which corresponds to an EP and consequently $M_c$ has a single eigenvector and a generalized eigenvector~\footnote{An EP is defined as a point in parameter space such that two eigen-values as well as the corresponding two eigen-vectors of a matrix coalesce (see~\cite{KatoEP} for details). Since a single eigen-vector exists at the EP, a basis can be obtained by adding a generalized eigen-vector. In our case (a $2 \times 2$ matrix) it means that such a vector $\hat V$ satisfies $(\lam - M)^2 \cdot \hat V = 0$ but $(\lam - M) \cdot \hat V \neq 0$. Notice that because $M_c$ is $2\times 2$, any vector not aligned with the eigenvector will satisfy this.}. The two are shown in Fig.~\ref{WM_Bloch_Vec_Fig}. The eigenvector is a standing wave with nodes of velocity (equivalently, pressure extrema) placed on the resistive sheets. Hence, one directly sees from the pressure jump condition \eqref{WM_jump} that this solution does not interact with the resistive sheets, which explains why its amplitude is not affected by it (Fig.~\ref{WM_Bloch_Vec_Fig} left panel). To build the second solution, let us consider a real-valued standing wave in one cell, with pressure nodes placed on the resistive sheets (so it is by construction linearly independent with the eigenvector). While moving across a sheet, the wave acquires an imaginary part proportional to the derivative of the real part (see \eq{WM_jump}). This imaginary part has its (pressure) extrema on the resistive sheets, and hence, it triggers no change of amplitude on the real part. Hence, we just build a solution made of a real part being a standing wave not affected by the resistive sheets, and an imaginary part having always the same increase in amplitude while passing across a sheet (Fig.~\ref{WM_Bloch_Vec_Fig} right panel). In other words, the amplitude varies linearly with the distance, and therefore we anticipate a transmission coefficient decreasing linearly with the size of the system. 

To conclude this section, we wish to underline two points. First, the occurrence of an EP of the transfer matrix is a consequence of two features: half the periodicity of the wave is the same as the distance between successive sheets and the symmetry of the wave is such that velocity nodes are aligned with the resistive sheets. Understanding this will allow us to manufacture EPs in more general setups, and therefore have similar anomalous transmission effects. This will be explored in Sec.~\ref{MeshandRes_Sec}. Second, when one has an EP, the eigenvector is necessarily accompanied by a generalized eigenvector with a linearly changing amplitude. Having this point in mind is central to understand the transmission properties of the medium. 

\subsection{Slab of layered medium}
\label{Slab_Scatt_Sec}
\begin{figure}[!ht]
\centering
\includegraphics[width=0.49\columnwidth]{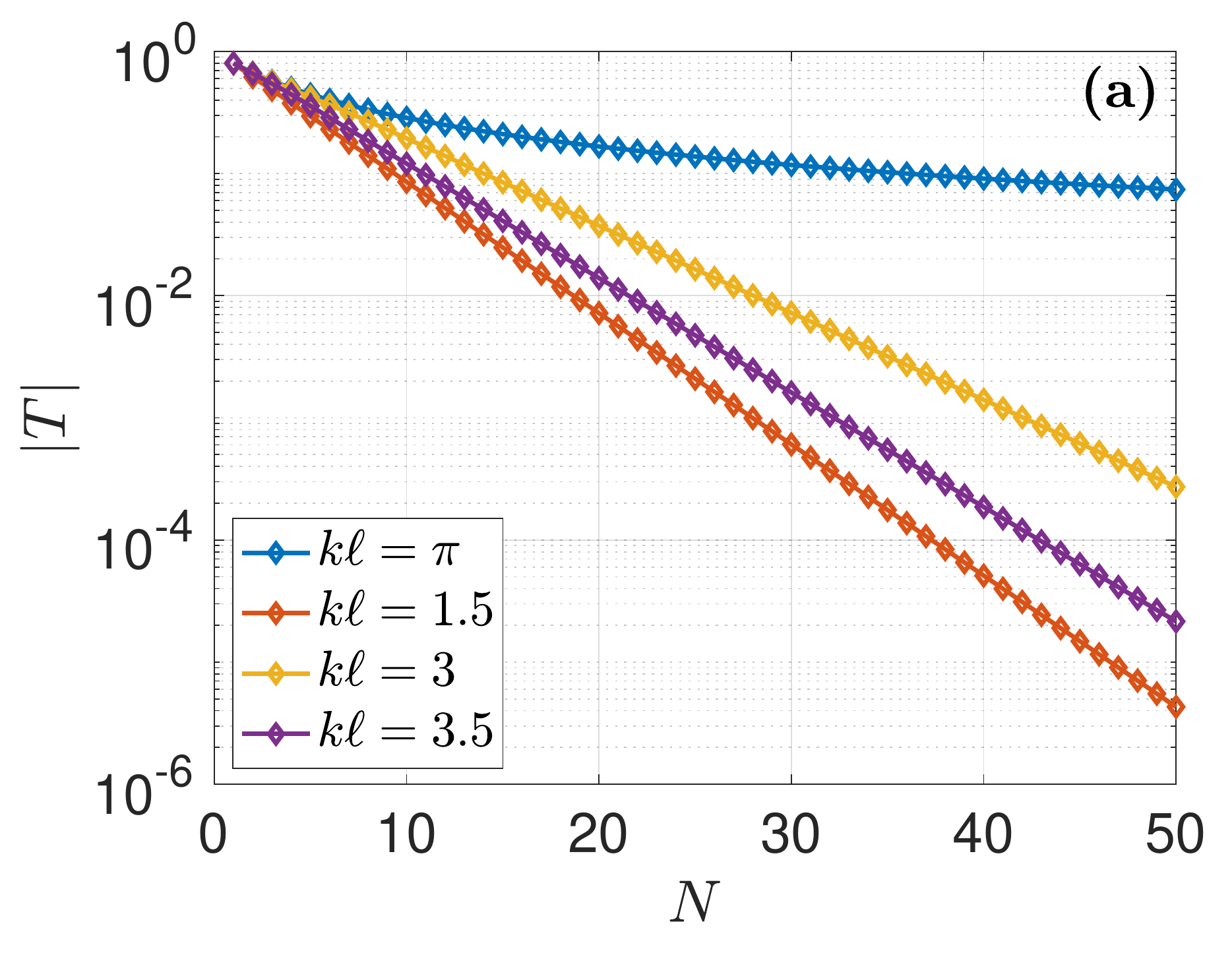}
\includegraphics[width=0.49\columnwidth]{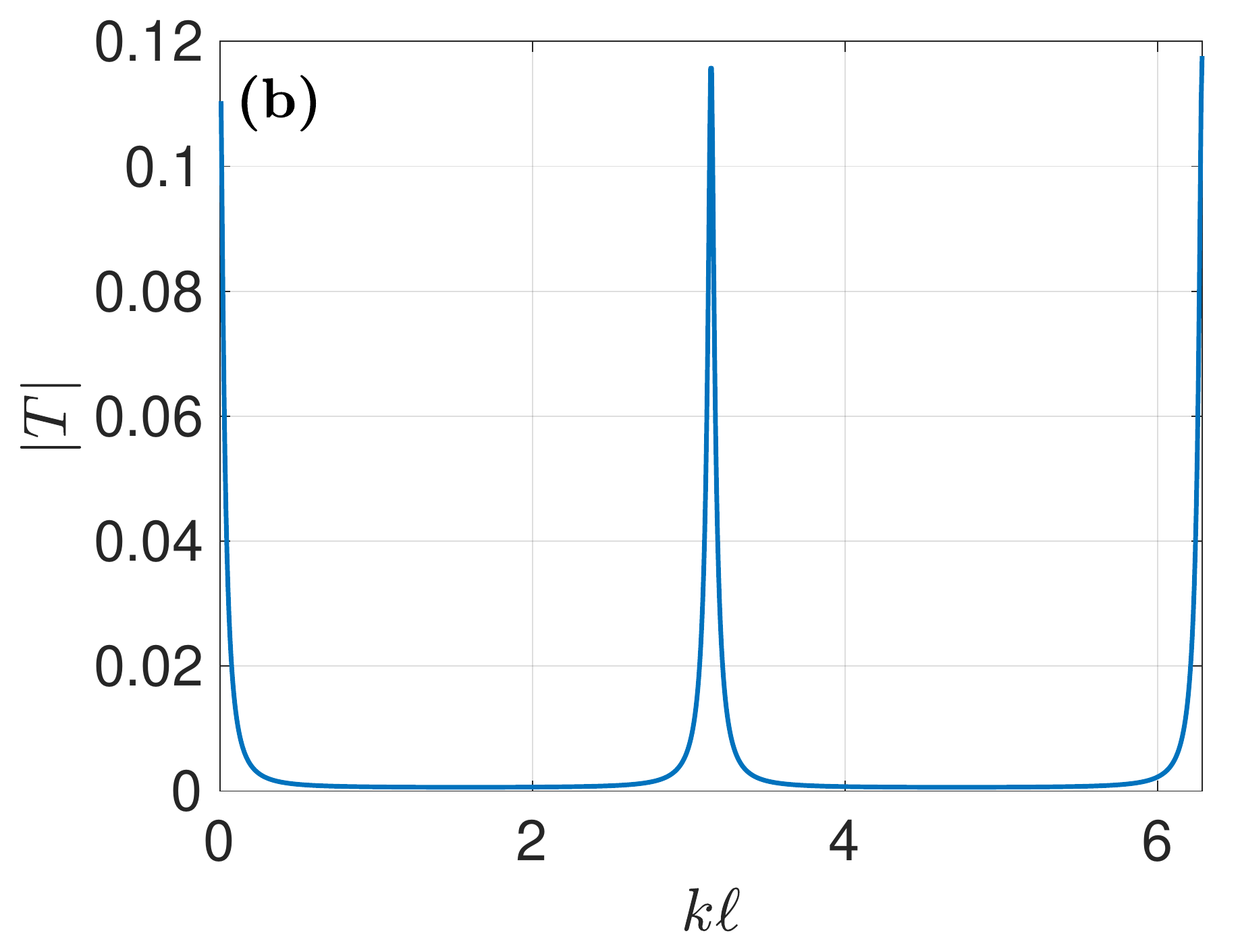}
\caption{(a) evolution of the transmission for different values of the frequency $k$ as a function of $N$. (b) scattering coefficients for $N=30$ resistive sheets with $\gam = 0.5$.}
\label{WM_Scatt_Fig} 
\end{figure}

We now discuss the scattering coefficients across a number $N$ of cells as in Fig.~\ref{WM_Schema_Fig}. We first notice that the total transfer matrix is given by the $N$-th power of $M_c$. We then use the same relation as \eqref{Mmat_def} to obtain the scattering coefficients: 
\bea \label{tot_Smat}
M_c^N &=& \bmat \vspace{8pt} T - \dfrac{R^2}{T} & \dfrac{R}{T} \\ - \dfrac{R}{T} &  \dfrac{1}{T} \emat.   
\eea
Because of the relative simplicity of the problem, all scattering coefficients can be obtained explicitly using the Chebyshev identity for $M_c^N$ (details are given in Appendix~\ref{Mmat_Smat_App}). The results are shown in Fig.~\ref{WM_Scatt_Fig}. We observe what was anticipated from the medium dispersion relation \eqref{Borr_DispRel}: The transmission decreases exponentially with the number $N$ of sheets, except at the Bragg frequency, where the decrease is instead linear. When fixing $N$ large and varying the frequency, this change of behavior manifests itself as a very sharp peak of transmission near $k \ell = \pi$. The more sheets are taken, the finer the peak is. 

To further characterize this transmission peak, we estimate the maximum and minimum transmission. When $N$ is large enough, the transmission outside of the Bragg frequency is a plateau near its minimum (see Fig.~\ref{WM_Scatt_Fig}). To estimate it, we can compute the transmission at the frequency $k\ell = \pi/2$ in the limit $N \gg 1$. This gives us 
\be \label{T_min}
\left| T_\m \right| \sim \left( \frac{2\sqrt{1+\gam^2/4}}{1+\sqrt{1+\gam^2/4}} \right) e^{-N \mathrm{argsh}\left(\gam/2\right)}. 
\ee
We point out that $|T|$ at $k\ell = \pi/2$ is not a strict minimum. Indeed, for moderate $N$, the amplitude of $|T|$ oscillates around its plateau value, and hence, depending on the parity of $N$, it will be either a local maximum or a local minimum. However, these oscillations are very small in the regime of interest, scaling as $O(e^{-N \mathrm{argsh}\left(\gam/2\right)})$.
When $\gam$ is small enough, the transmission coefficient $T_\m$ takes the simpler asymptotic form
\bsub \label{T_min_as} \bea
\left| T_\m \right| &\sim& e^{-N \gam/2} , \\
&\sim& \left(1+\frac{\gam}{2}\right)^{-N} . 
\eea \esub
On the second line we recognize the product of $N$ times the transmission of a single sheet $t_1$. In other words, outside of the Bragg frequency, the effects of the sheets multiply. This is strictly speaking true for $\gam \ll 1$, but we see from \eqref{T_min} that the behavior is similar also for large $\gam$. 

The maximum transmission is reached at the Bragg frequency, and takes the simple form 
\be \label{T_max}
|T_\M| = \frac{1}{1+N \gam/2}. 
\ee
This can be obtained directly from the general expression of $M_c^N$ (see Appendix~\ref{Mmat_Smat_App} \eq{Borr_General_Scatt}), or by noticing that at the Bragg frequency, $M_c = - I_2 + \frac{\gam}2 A$ with $A^2 = 0$, and hence, $M_c^N = (-1)^N (I_2 - \frac{N\gam}2 A)$. We notice that this is the transmission of a single sheet of resistance $N\gam$. In other words, at the Bragg frequencies, the effects of the different sheets add up. 

\begin{figure}[!ht]
\centering
\includegraphics[width=0.49\columnwidth]{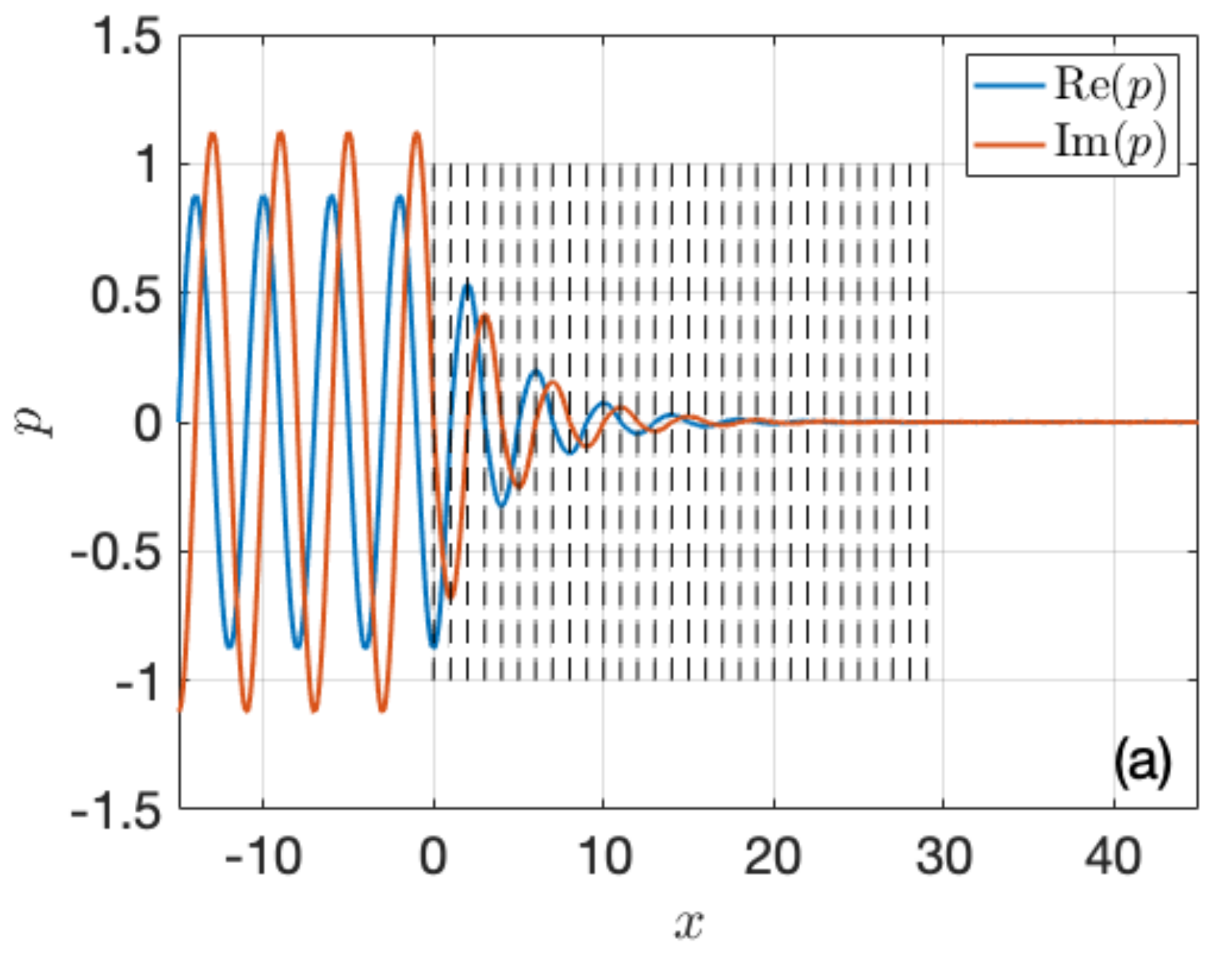}
\includegraphics[width=0.49\columnwidth]{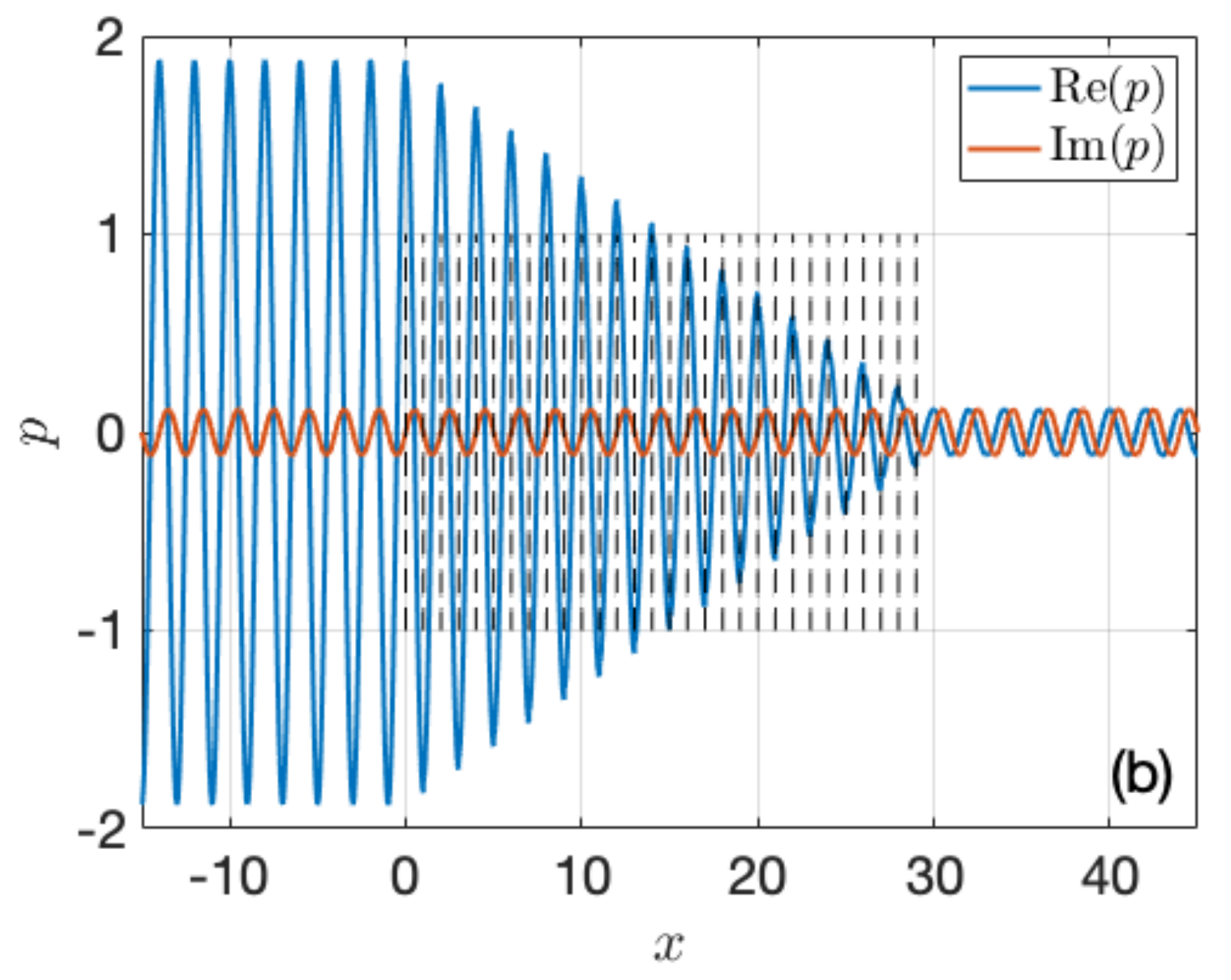}
\caption{Scattering solutions on a slab of $N=30$ resistive sheets (marked by dashed black lines) with $\gam = 0.5$. The locations of the resistive sheets are shown as black dashed lines.  (a) At $k\ell = \pi/2$ with the classical exponential decrease. (b) At $k\ell = \pi$ where there is the EP associated with a linear decrease.} 
\label{WM_Modes_Fig} 
\end{figure}

To further understand the two characteristic behaviors of the transmission, we represented the scattering solutions in Fig.~\ref{WM_Modes_Fig}. Outside the Bragg frequency, the amplitude of the wave decreases exponentially while the wave propagate in the material, as expected within a dissipative medium. On the contrary, the structure of the solution at the Bragg frequency is rather unusual, with part of the wave ($\Im(p)$ in Fig.~\ref{WM_Modes_Fig}) being seemingly unaffected by the sheets, and another part ($\Re(p)$ in Fig.~\ref{WM_Modes_Fig}) with a linearly decreasing amplitude. This can be readily understood by considering the two linearly independent solutions of the periodic medium as described in Sec.~\ref{WM_Bloch_Sec}. Since the scattering solution we consider is purely right moving at the end of the slab, it cannot be proportional to the eigenvector of $M_c$, which is a standing wave. Hence it is a combination of the eigenvector and generalized eigenvector, which explains that part of the wave is unaffected by the resistive sheets, while the other undergoes an amplitude decrease linear with the distance. 

\subsection{Width of the anomalous transmission peak}
\label{Finesse_Sec}
A remarkable property of the anomalous transmission observed in Fig.~\ref{WM_Modes_Fig} is that the peak becomes increasingly narrow as $N$ is taken larger. To see this, we analyze the transmission coefficient in the vicinity of the EP. This is legitimate in the limit of a large number of resistive sheets $N \gg 1$, since the peak becomes finer in that limit. For this we define 
\bsub \bea
k \ell &=& \pi + \delta , \\
q \ell &=& \pi + \sigma , 
\eea \esub
and assume small $\delta$ and $\sigma$. The dispersion relation \eqref{Borr_DispRel} then reduces to 
\be \label{NearPeak_DispRel}
\sigma^2 = i \gam \delta. 
\ee
We now look at the vicinity of the peak by assuming $|\sigma| \ll 1$ and $N \gg 1$, but without any restriction on $N\sigma$. In this limit, the transmission coefficient simplifies into 
\be
|T|^2 \sim \frac{4 |\sigma|^2}{\gam^2 |\sin(N \sigma)|^2} \sim \frac{4 |\delta|}{\gam \sin^2 \left(N \sqrt{\gam |\delta|/2} \right)+ \gam \sinh^2 \left(N \sqrt{\gam |\delta|/2}\right)}. 
\ee
We now define the width as the value $\delta = \pm \ell \Delta k$ such that $|T| = T_\M/2$. From the above expression we obtain 
\be \label{Peak_Finesse}
\ell \Delta k = \frac{2\zeta_0^2}{\gam N^2}, 
\ee
where we introduced the numerical constant $\zeta_0 \approx 2.1425$ defined as the unique nonzero solution of $4\zeta_0^2 = \sin^2(\zeta_0)+\sinh^2(\zeta_0)$. From \eq{Peak_Finesse}, we see that the peak becomes rapidly finer as $N$ increases. It is also interesting to notice that the maximum and minimum transmission are both essentially governed by $N\gam$ (see Eqs.~\eqref{T_min_as} and \eqref{T_max}), so one can maintain those fixed while increasing the peak width by adding more cells.

\subsection{Effect of thickness}
\label{Thick_Borr_Sec}
\begin{figure}[!ht]
\centering
\includegraphics[width=0.49\columnwidth]{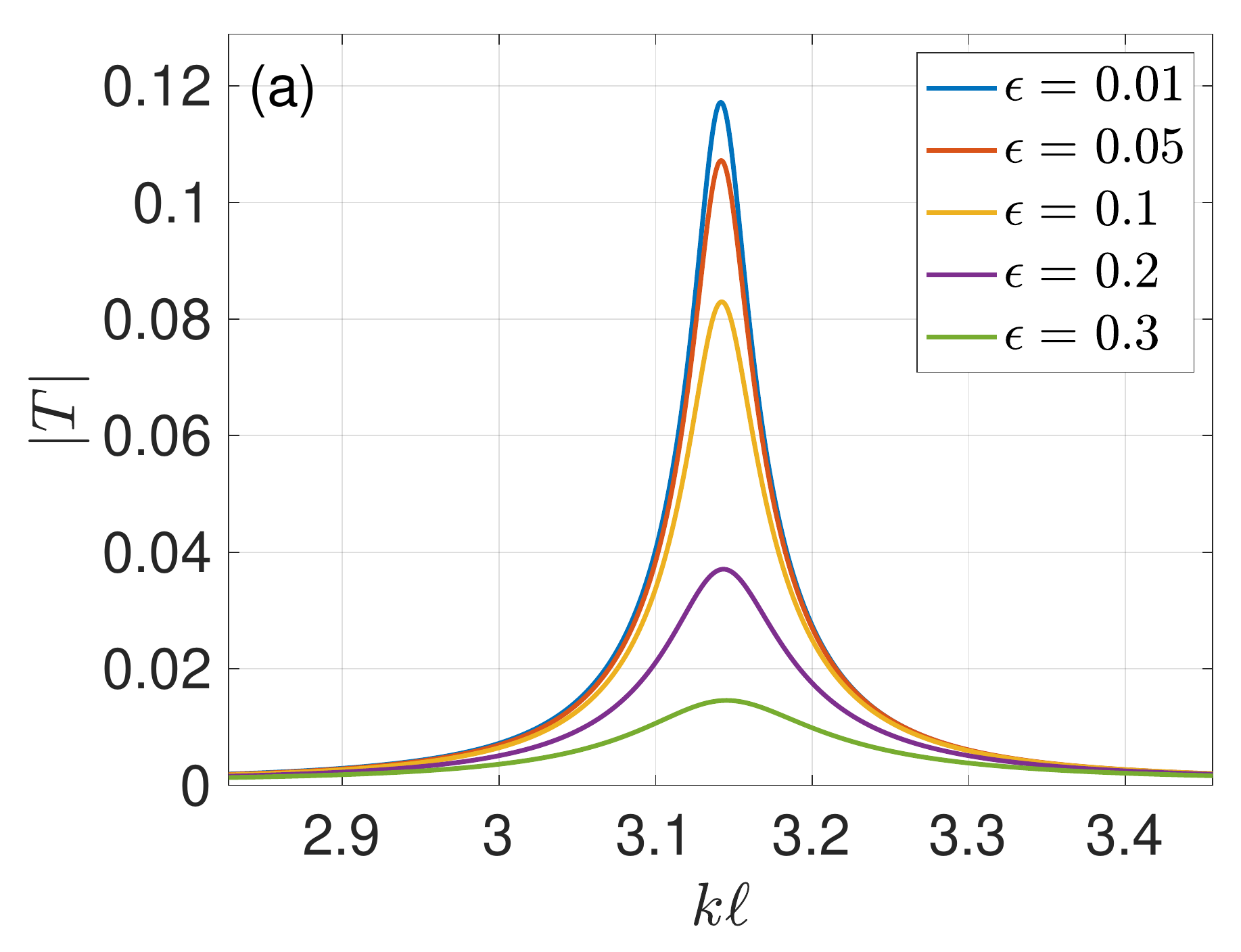}
\includegraphics[width=0.49\columnwidth]{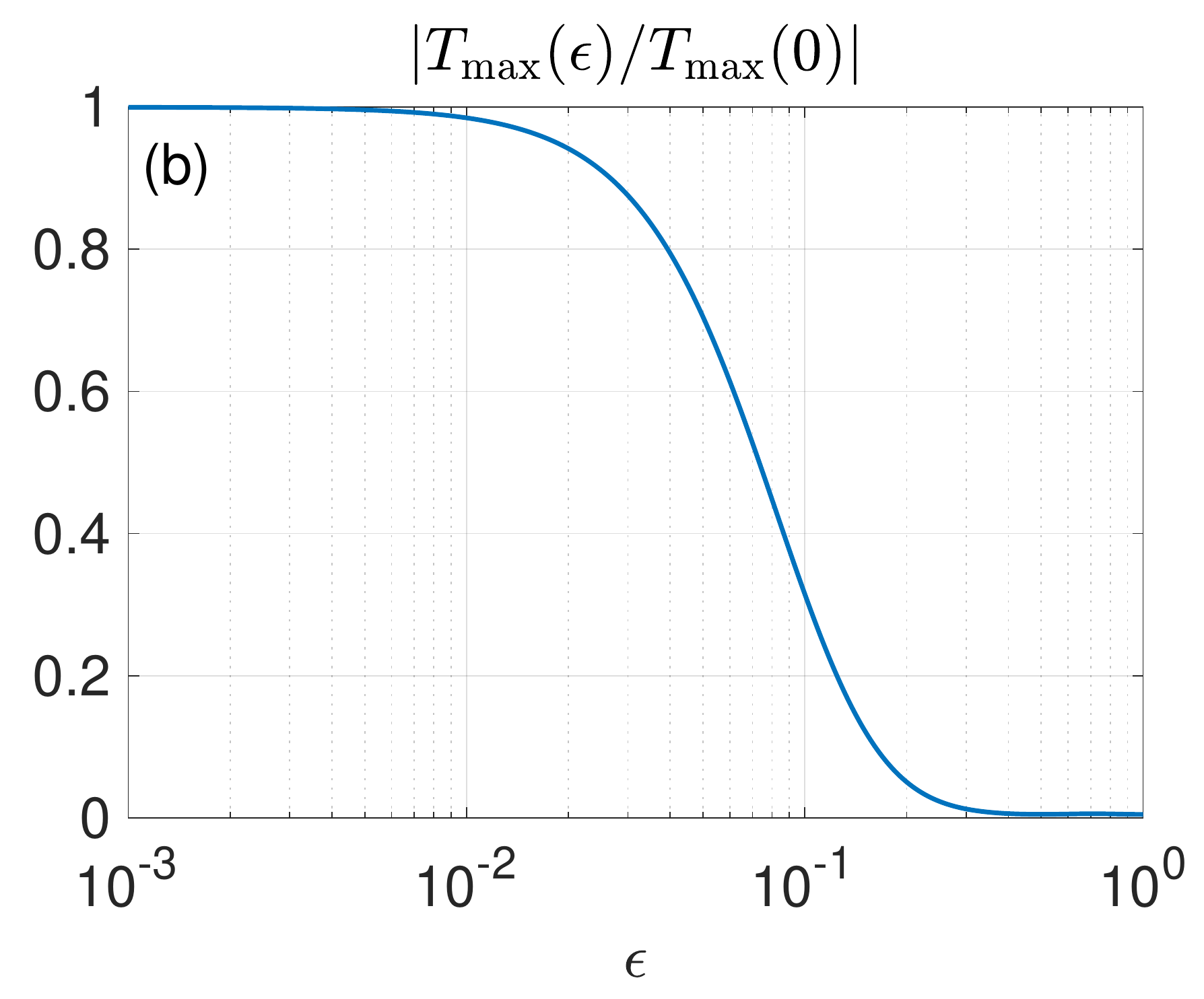}
\caption{(a) transmission coefficient against the frequency, near the Bragg frequency. We vary the thickness fraction $\epsilon=\ell_0/\ell$ while maintaining $\gam = 0.5$ fixed. We used $N=30$ cells. (b) maximum transmission (for $k\ell = \pi$) as a function of the thickness fraction $\epsilon=\ell_0/\ell$.}
\label{2D_plot_Fig} 
\end{figure}

We now analyze a second key property of the anomalous transmission peak of Fig.~\ref{WM_Scatt_Fig}: the smaller the sheet thicknesses are compared to the wavelength, the stronger the transmission peak is. To illustrate this, we consider a dissipative slab made of open horizontal pores of length $\ell_0$. This corresponds to a simple model of porous material~\cite{Allard}. The net effect of the slab can be described with an effective (complex) dimensionless density $\rho_\eff(k)$ of fixed (independent of $\ell_0$) total resistance $\gam$
\be \label{LF_porous}
\rho_\eff = \sqrt{1+\frac{i \gam}{k \ell_0}}. 
\ee
We also assume that the effective bulk modulus is the same as in air (this amounts to neglecting visco-thermal effects in the material, which is legitimate at low frequencies). Using \eqref{LF_porous}, the Helmholtz equation in the material reads 
\be \label{Helm_eq_porous}
\p_x^2 p + k^2 \rho_\eff p = 0. 
\ee
We now put this slab inside a cell of length $\ell$, similarly to Fig.~\ref{WM_Schema_Fig}. Imposing continuity of pressure and velocity at the interfaces allows us the extract the transfer matrix of the cell. From this we obtain the transmission and reflection coefficients of a single cell: 
\bsub \label{Porous_Mmat} \bea 
t_1 &=& \frac{2\rho_\eff}{2\rho_\eff \exp(i ((\rho_\eff -1) \ell_0 + \ell) k) - i (\rho_\eff-1)^2 \sin(\rho_\eff \ell_0 k)e^{-i (\ell - \ell_0) k}} , \\
r_{1} &=& \frac{i (\rho_\eff^2-1) \sin(\rho_\eff \ell_0 k)}{2\rho_\eff \exp(i ((\rho_\eff -1) \ell_0 + \ell) k) - i (\rho_\eff-1)^2 \sin(\rho_\eff \ell_0 k)e^{-i (\ell - \ell_0) k}} , 
\eea \esub
A direct verification shows that the limit of \eq{Porous_Mmat} for $\ell_0 \to 0$ gives back the case of a resistive sheet as described by \eq{Mmat_cell}. Moreover, we verify that the porous slabs have the same total resistance at zero frequency for all $\ell_0$. This is done by imposing that the scattering coefficients in the zero frequency limit verify  
\be
\left| \frac{r_1}{t_1}\right|_{k \to 0}^2 = \gam^2/4, 
\ee
independently of $\ell_0$. When looking at the Bloch wave solutions, the EP at $k\ell = \pi$ is replaced by an avoided crossing (see Fig.~\ref{WM_Bloch_Fig} (b)). We therefore anticipate that the transmission peak will be reduced. In Fig.~\ref{2D_plot_Fig}, we show the transmission coefficient across a periodic piece of such porous slabs. We observe that the plateau of transmission outside the Bragg frequency has the same value for all $\ell_0$, meaning that this value only depends on the total resistance $\gam$. On the contrary, the maximum transmission at the anomalous peak increases significantly as the dissipative slab becomes of sub-wavelength size. As we see in Fig.~\ref{WM_Bloch_Fig}, the optimal transmission is reached for $\ell_0/\ell \lesssim 1\%$. 

\section{Experimental results}
\label{Exp_Sec}

\begin{figure}[!ht]
\centering
\includegraphics[height=0.21\textheight]{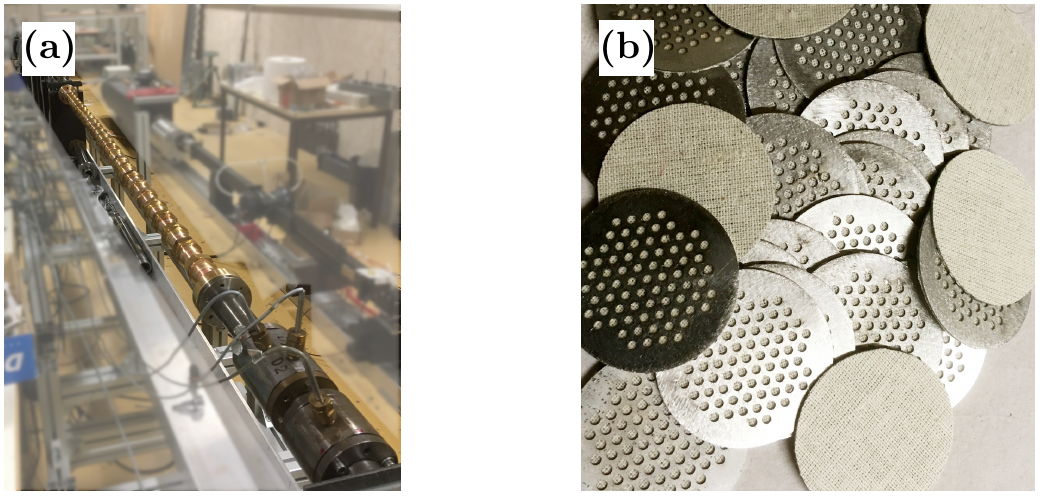}
\caption{Pictures of the experimental setup. (a) cylindrical duct with speakers and microphones on both ends. (b) resistive sheets made of fine meshed fabric mounted on a perforated metallic plate. The radius of both the duct and the resistive sheets is $3 \mathrm{cm}$.}
\label{ExpPic_Fig} 
\end{figure}

For the experimental characterization of the Borrmann effect with resistive sheets, we used a rigid cylindrical duct of radius $3 \mathrm{cm}$, with two measurement sections on each end. Each of them comprises two microphones and an acoustic source, allowing to extract reflection and transmission coefficients. A more detailed description of the experimental setup can be found in~\cite{Testud09} (see also Fig.~\ref{ExpPic_Fig}). Inside the duct, we put resistive sheets made of a fine meshed fabric mounted on a metallic perforated plate. All sheets are separated by a distance $\ell = 7 \pm 0.2 \mathrm{cm}$. The thickness of the resistive sheets is below $0.5 \mathrm{mm}$, so we are in the subwavelength regime (see Sec.~\ref{Thick_Borr_Sec}). The aim of the metallic plate is to reduce the elastic motion of the sheets. In practice there is a small added mass to the sheet. This is modelled by replacing $\gam$ in the jump condition \eqref{WM_jump} by a complex (adimensional) impedance $z = \gam - i \mu k \ell$. $\gam$ is the resistance and $\mu$ the added mass. Both coefficients are real and adimensional. We also take into account the losses due to friction on the wall of the waveguide (see for instance Sec. 4.5.3 in~\cite{Rienstra01}). 

\begin{figure}[!ht]
\centering
\includegraphics[width=0.49\columnwidth]{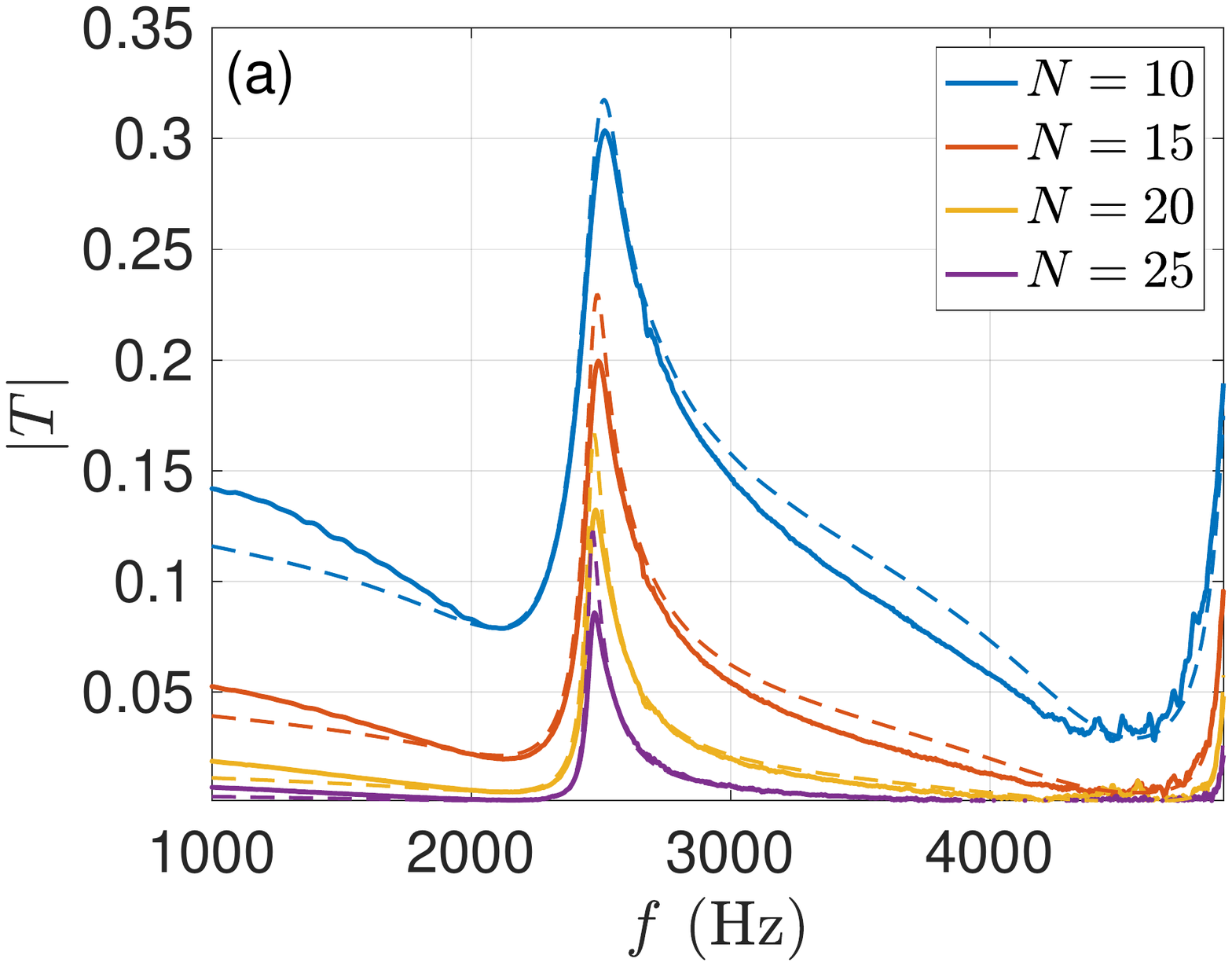}
\includegraphics[width=0.49\columnwidth]{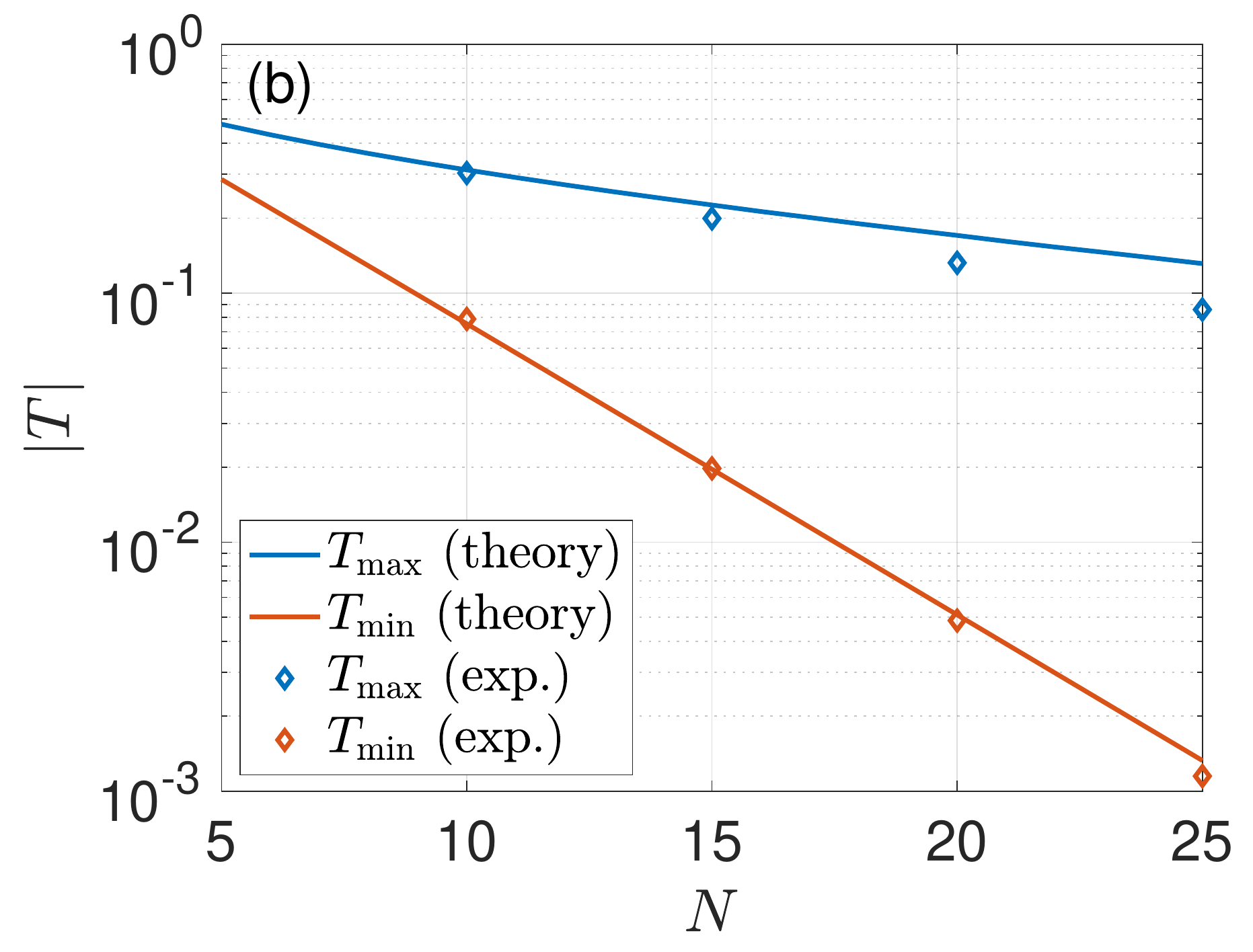}
\caption{(a) transmission coefficient against the frequency for different number of resistive sheets. Continuous lines are experimental data and dashed lines are theory. (b) maximum transmission and minimum transmission (taken for lower frequencies than Bragg) as a function of the number of sheets $N$. Diamonds are experimental data and continuous lines are theory.}
\label{Exp_Fig} 
\end{figure}

We conducted the scattering experiment for $N=10$, $15$, $20$, and $25$ resistive sheets. The results are shown in Fig.~\ref{Exp_Fig}. We clearly see the anomalous transmission at the Bragg frequency. We also see that the peak becomes narrower as $N$ is increased. The value of the impedance is determined by comparing the theory with experimental data using a least square method. We found $\gam = 0.46 \pm 0.03$ and $\mu = 0.11 \pm 0.01$. The value of the resistance was measured using an independent method, and coincides within error bars. Additionally, we notice that the transmission peak becomes asymmetric compared to Fig.~\ref{WM_Scatt_Fig}. This is due to the non-zero imaginary part of the impedance and can be understood in the following way: when $z$ is purely imaginary ($\Im(z) > 0$), a gap opens just below the Bragg frequency; hence, transport at frequencies below Bragg are inhibited, while it is facilitated for frequencies above Bragg, which is what is observed in Fig.~\ref{Exp_Fig}. We also compared the scaling of the maximum and minimum transmission as the number of sheets $N$ is increased (Fig.~\ref{Exp_Fig} right panel). We can see that the decrease in transmission is slower at the Bragg frequency.

\section{Anomalous transmission in more general periodic media}
\label{MeshandRes_Sec}

In this section, we show how to obtain a similar anomalous transmission by placing resistive sheets inside a (lossless) structured periodic medium. This has two main interests. First, by tuning the base medium, we can change the frequency at which the anomalous transmission peak occurs. In particular, this allows us to obtain a peak at a wavelength significantly larger than the size of the unit cell. Second, the value of the maximum transmission is in general increased compared to an array of resistive sheets alone (that is compared to \eq{T_max}). 

\subsection{Manufacturing exceptional points}

\begin{figure}[!ht]
\centering
\includegraphics[width=0.7\columnwidth]{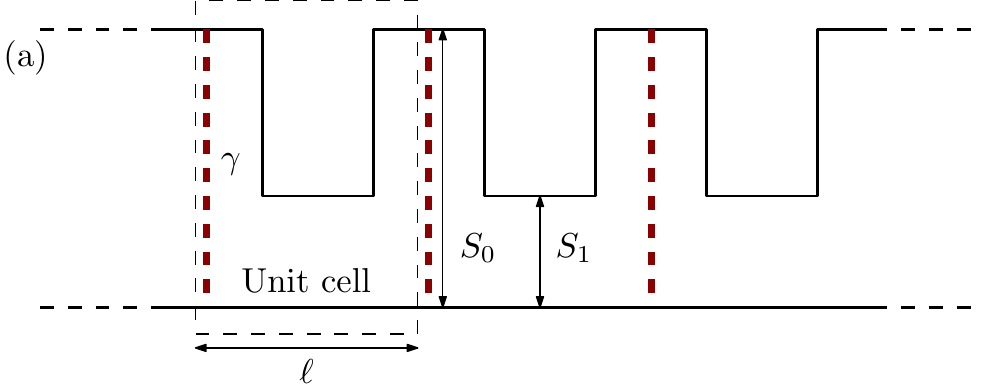}

\includegraphics[width=0.7\columnwidth]{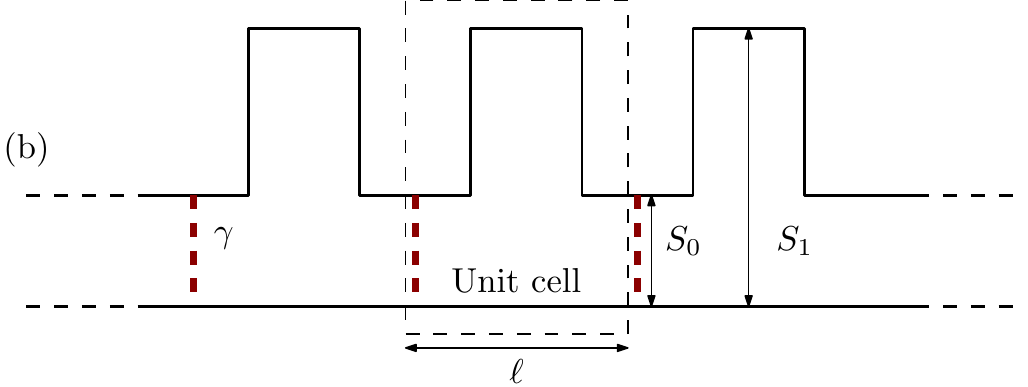}
\caption{Design of a configuration with resistive sheets and resonators in series. (a) Resistive sheets placed in the middle of the large cross-section parts. (b) Resistive sheets placed in the middle of the small cross-section parts.} 
\label{ResAndMesh_series_Fig} 
\end{figure}

For our base medium we use a waveguide with periodic changes of cross section: a first tube of cross section $S_0$ and length $\ell_0$ followed by one of cross section $S_1$ and length $\ell_1$, and repeated on a spatial period $\ell = \ell_0+\ell_1$ (see Fig.~\ref{ResAndMesh_series_Fig}). This corresponds to an array of Helmholtz resonators in series, the small cross section playing the role of the neck and the large one being the cavity.  

\begin{figure}[!ht]
\centering
\includegraphics[width=0.49\columnwidth]{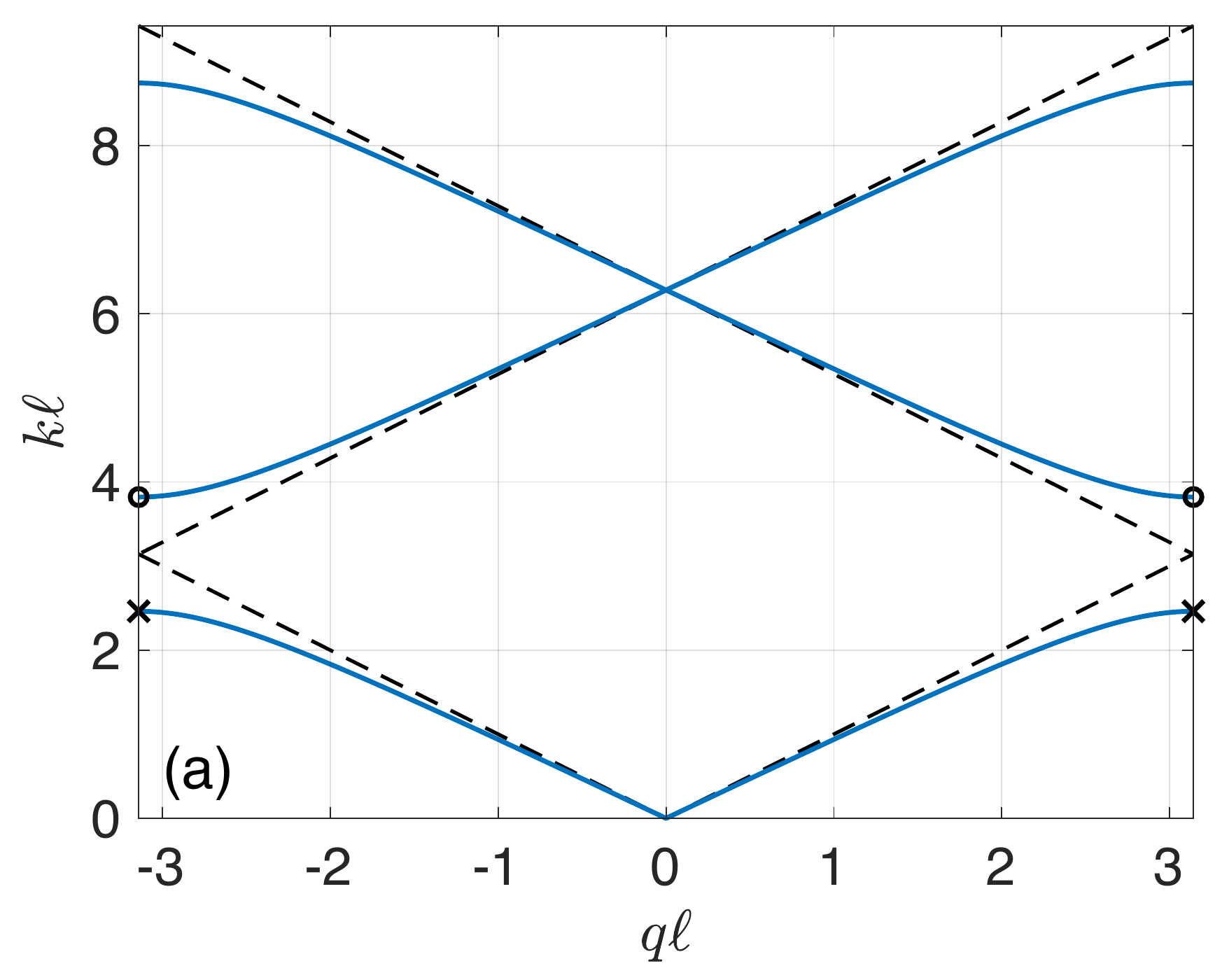}
\includegraphics[width=0.49\columnwidth]{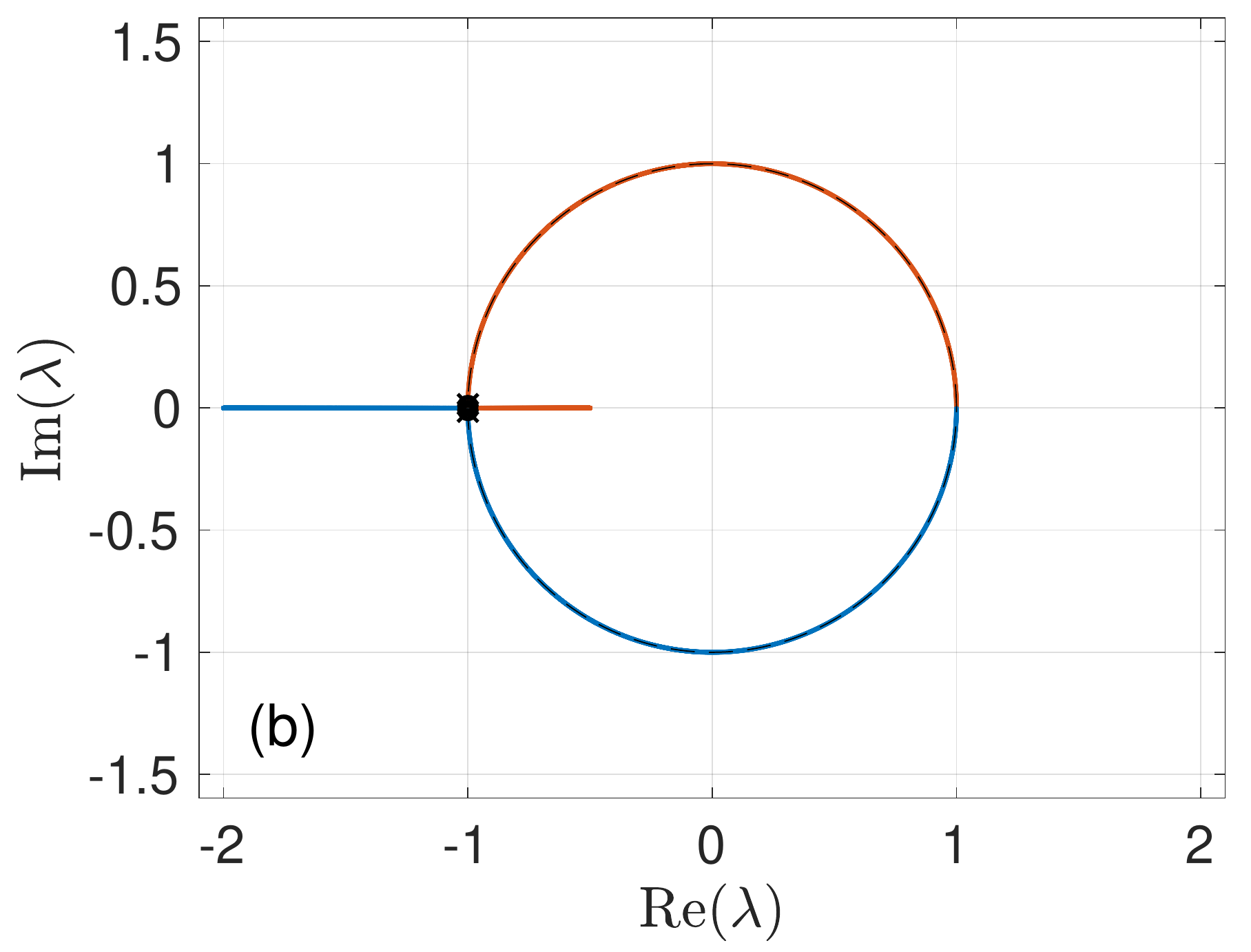}

\includegraphics[width=0.49\columnwidth]{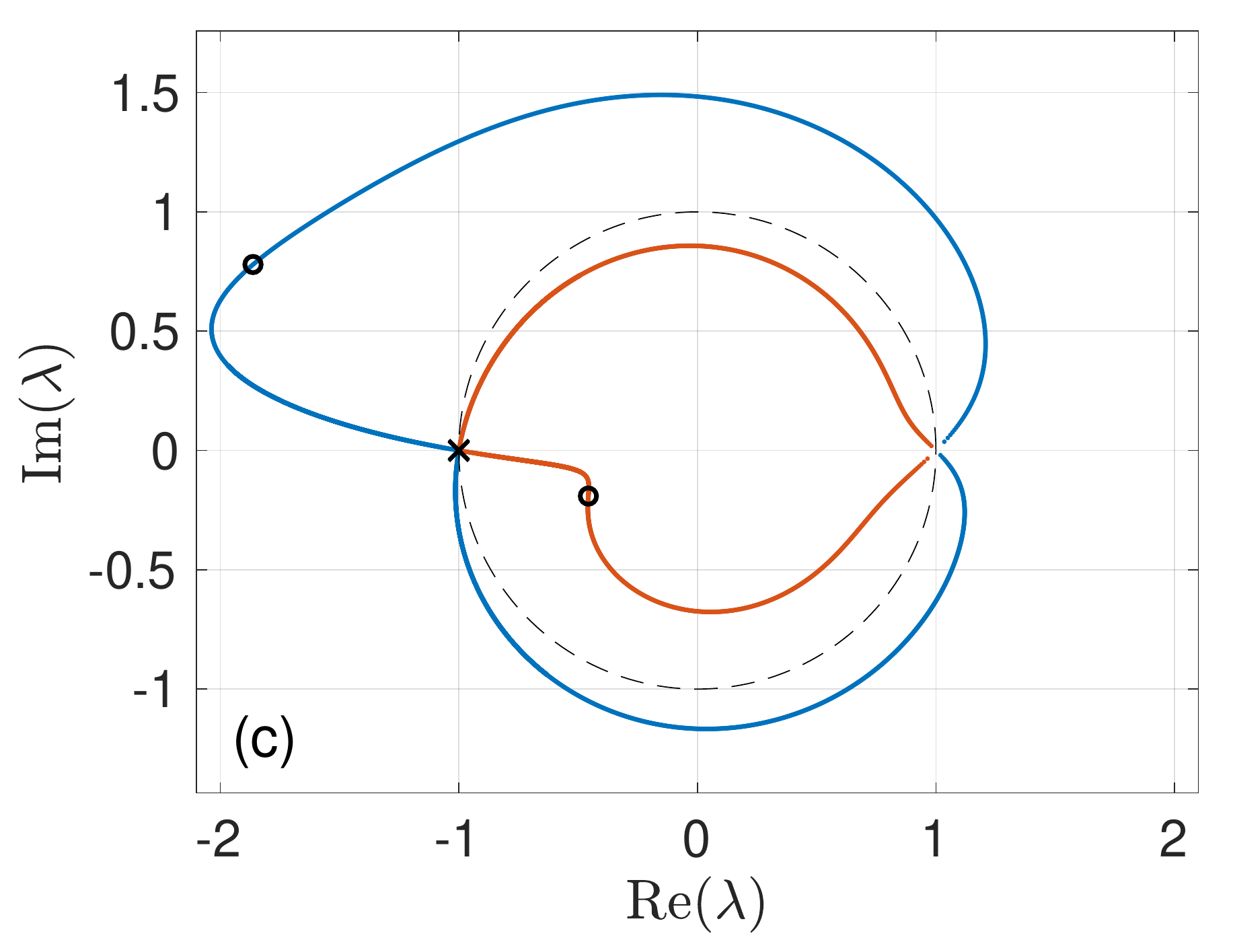}
\includegraphics[width=0.49\columnwidth]{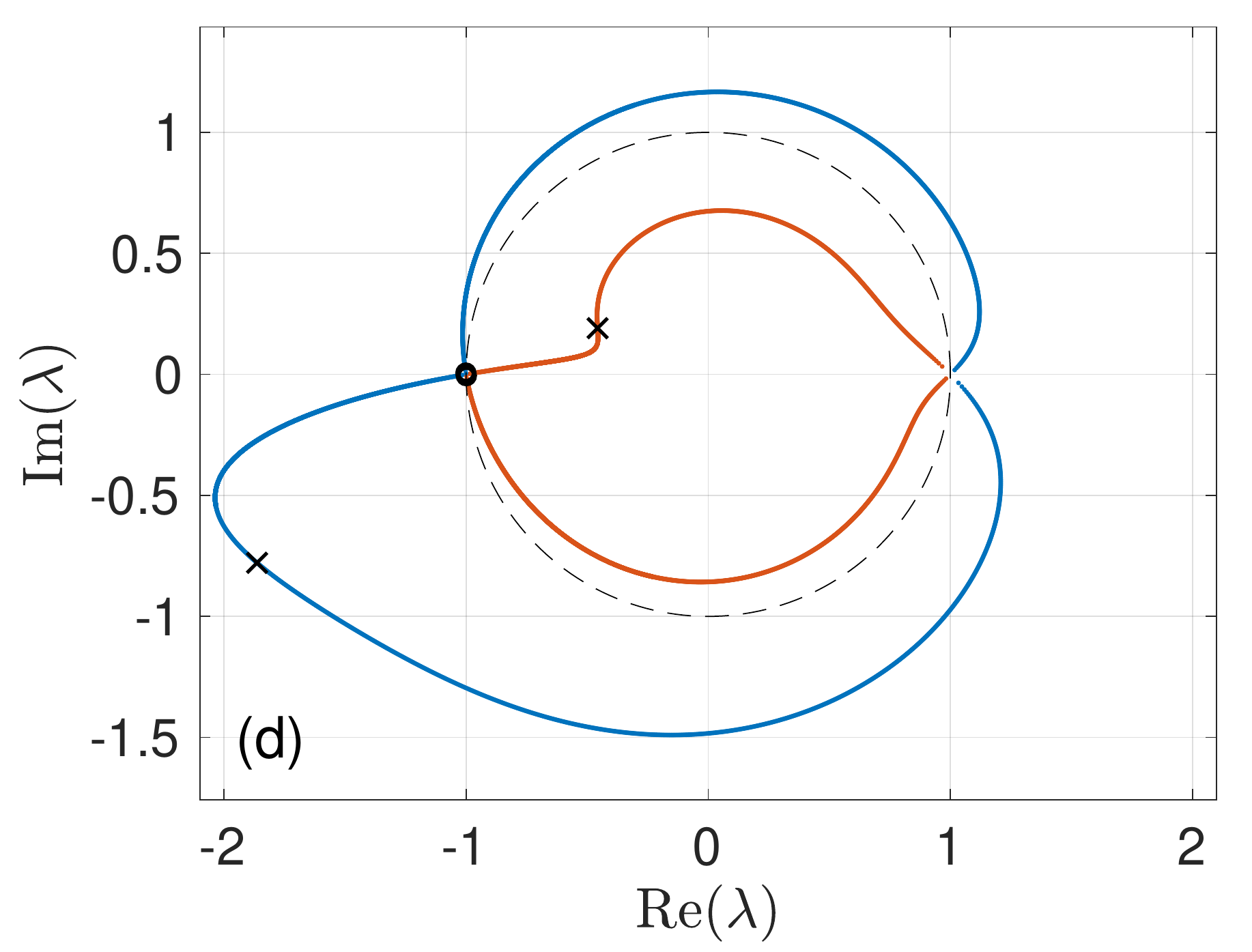}
\caption{Dispersion relation and eigenvalues $\lam=e^{\pm i q \ell}$ of the transfer matrix. We used $\nu = 2$, $\ell_1/\ell=0.5$, and $\gam=0.5$ for (b) and (c). (a) Dispersion relation of the base medium ($\gam=0$). The crosses (resp. circles) shows the values of $k$ at the lower edge (resp. upper) of the gap (also shown in (b-d)). (b-d) Eigenvalues of the transfer matrix for the setup of Fig.~\ref{ResAndMesh_series_Fig} and $k \ell$ from 0 to $2\pi$: (b) without resistive sheets, (c) resistive sheets in the large cross section tube (Fig.~\ref{ResAndMesh_series_Fig} (a)), and (d) resistive sheets in the small cross section tube (Fig.~\ref{ResAndMesh_series_Fig} (b)).}
\label{Bloch_ResAndMesh_series_Fig} 
\end{figure}

To simplify, we consider that the transverse lengths of the guide are much smaller than the cell length (i.e. $S_{0,1} \ll \ell^2$) and hence the typical wavelengths. In this limit, evanescent modes excited near each cross-section change can be safely neglected. Therefore, acoustic propagation reduces to that of the plane mode. This means that the system is effectively one-dimensional and can be describe using the same transfer matrix formalism as before. At cross section changes, pressure and debit are continuous. Using these conditions, we obtain the transfer matrix (see Appendix~\ref{Mmat_Smat_App}), which leads to the dispersion relation 
\be
2\cos(q \ell) = \dfrac{(1+\nu)^2}{2\nu} \cos(k \ell) - \dfrac{(1-\nu)^2}{2\nu} \cos(k (\ell_0 - \ell_1)) , 
\ee
where $\nu = S_0/S_1$ is the cross section ratio. This dispersion relation presents bands and gaps, as shown in Fig.~\ref{Bloch_ResAndMesh_series_Fig}~(a). When looking at the eigenvalues $e^{\pm i q \ell}$ of the transfer matrix, each edge of a band gap is an EP (see Fig.~\ref{Bloch_ResAndMesh_series_Fig}~(b)). Upon adding dissipation, eigenvalues are pushed away from the unit circle, and EPs are avoided. This encodes the fact that waves cannot propagate without attenuation. The idea is to place resistive sheets in the base medium such that a specific EP stays exceptional. 

To do so we use mirror symmetric cells, with either the large cross section tube or the small cross section tube in the middle (see Fig.~\ref{ResAndMesh_series_Fig}). At an edge of a band gap, the (unique) eigenvector of the transfer matrix is either symmetric or anti-symmetric~\cite{Xiao14}. This corresponds to a standing waves with pressure nodes either in the center (symmetric vector) of the cell or on the edges (anti-symmetric vector). Now, by adding resistive sheets at the location of the velocity nodes (pressure extrema), the same argument as in Sec.~\ref{WM_Bloch_Sec} shows that this standing wave is insensitive to the presence of the sheets. Hence, one obtains an EP for the transfer matrix, with similar properties as in Sec.~\ref{WM_Bloch_Sec} (a detailed proof of this statement is given in Appendix~\ref{General_EP_App}). This is shown in Fig.~\ref{Bloch_ResAndMesh_series_Fig} for our configuration with an array of resonators in series and resistive sheets (configuration of Fig.~\ref{ResAndMesh_series_Fig}). The wave at the lower edge of the first gap (marked as crosses in Fig.~\ref{Bloch_ResAndMesh_series_Fig}) has its velocity nodes in the center of the large cross section tube, while that of the upper edge has its velocity nodes in the center of the small cross section tube.

\subsection{Transmission peak}

\begin{figure}[!ht]
\centering
\includegraphics[width=0.49\columnwidth]{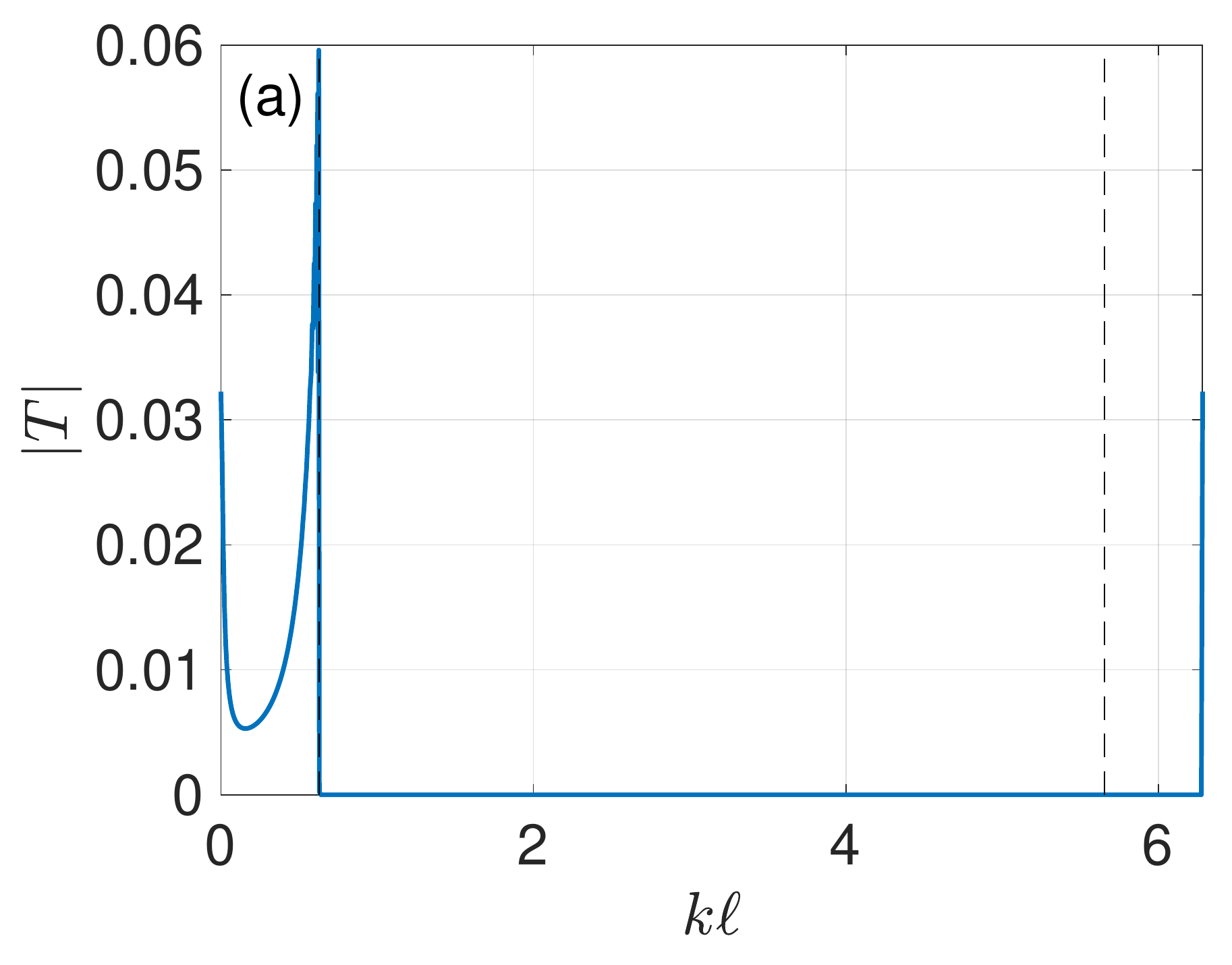}
\includegraphics[width=0.49\columnwidth]{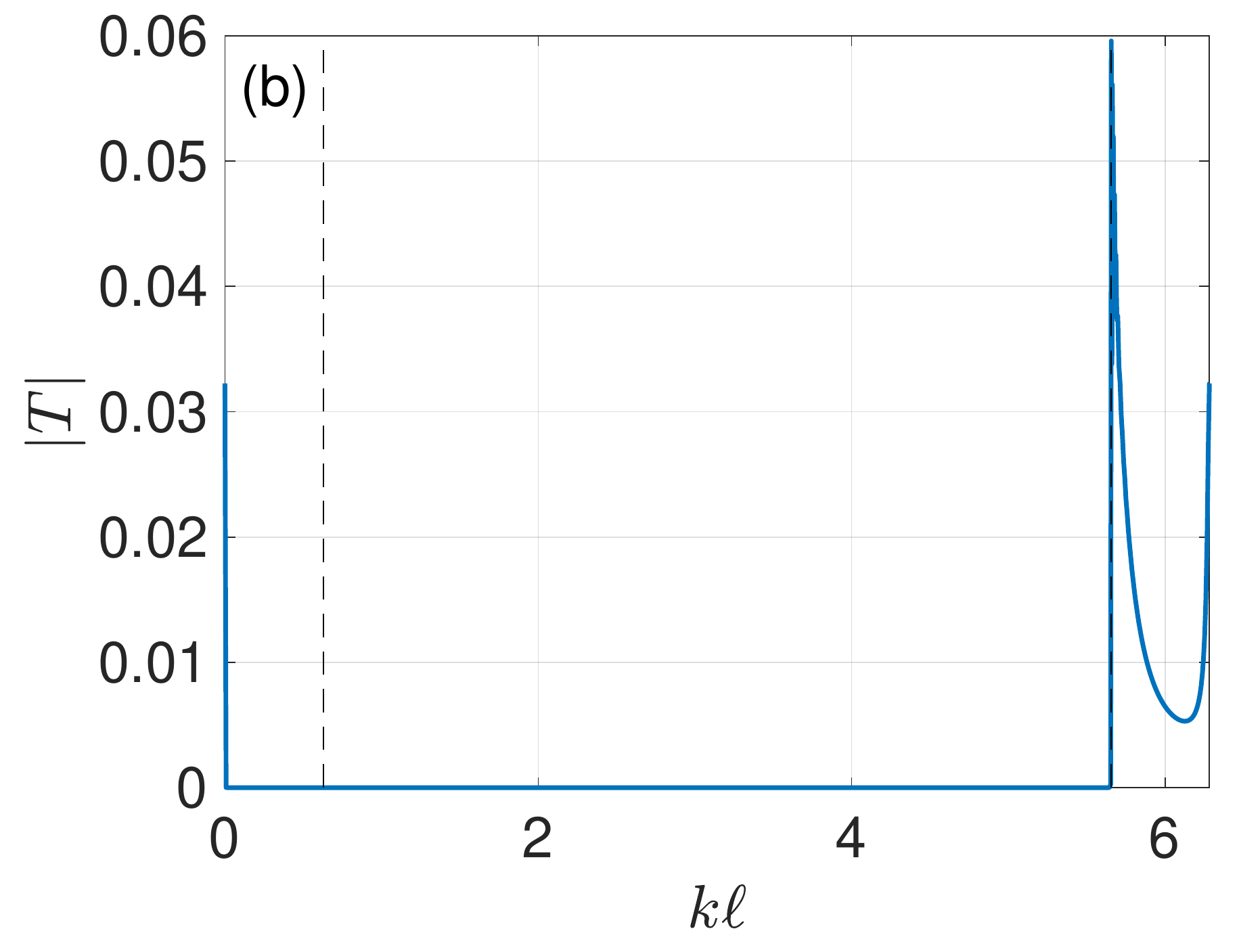}
\caption{Transmission coefficients as a function of the frequency for the configurations of Fig.~\ref{ResAndMesh_series_Fig}. The upper and lower edge of the first gap are indicated by dashed lines. We used $N=30$ unit cells, $\nu = 40$, $\ell_1/\ell = 0.5$ and $\gam = 2$. For this cross-section ratio, the wavelength of the lower edge is ten times smaller than the size of the unit cell. (a) We placed the resistive sheets in the center of the large cross section tube (Fig.~\ref{ResAndMesh_series_Fig}~(a)). (b) We placed the resistive sheets in the center of the small cross section tube (Fig.~\ref{ResAndMesh_series_Fig}~(b)).}
\label{Borr_ResAndMesh_series_Fig} 
\end{figure}

Once an EP is obtained, the transmission properties of the medium (with resistive sheets) follow as in Sec.~\ref{WM_Bloch_Sec}: generic frequencies (not associated with an EP of the transfer matrix) are attenuated exponentially with the distance either due to dissipation or because they are in the gap and hence are evanescent. On the contrary, at the EP, the transmission is decreasing linearly with the distance due to a combination of a standing wave insensitive to the dissipation and a generalized eigenvector with linearly changing amplitude. The transmission as a function of the frequency is shown in Fig.~\ref{Borr_ResAndMesh_series_Fig}. We see that depending on the location of the resistive sheets in the base medium, the transmission peak is found on one edge or the other of the band gap. 

\begin{figure}[!ht]
\centering
\includegraphics[width=0.49\columnwidth]{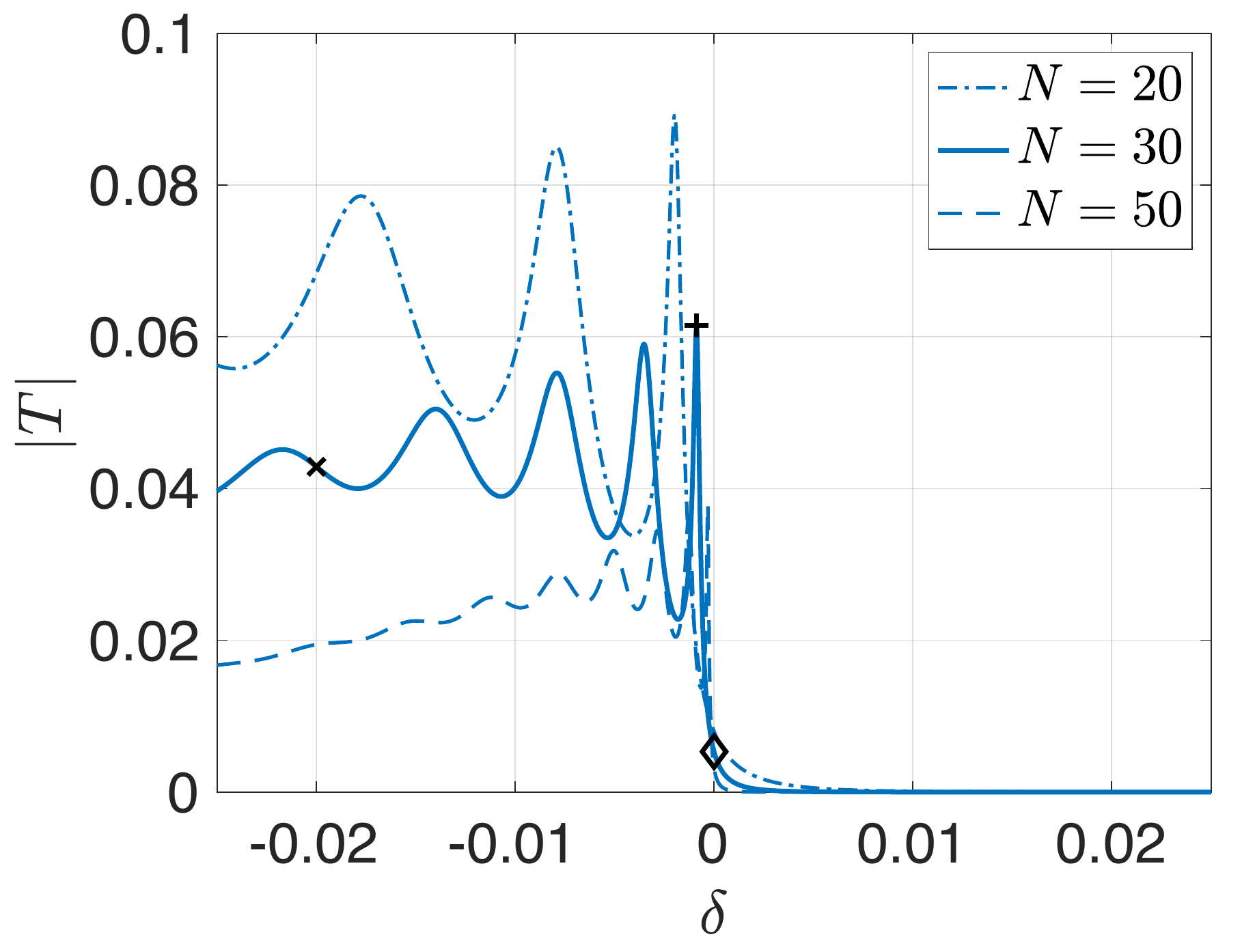}
\includegraphics[width=0.49\columnwidth]{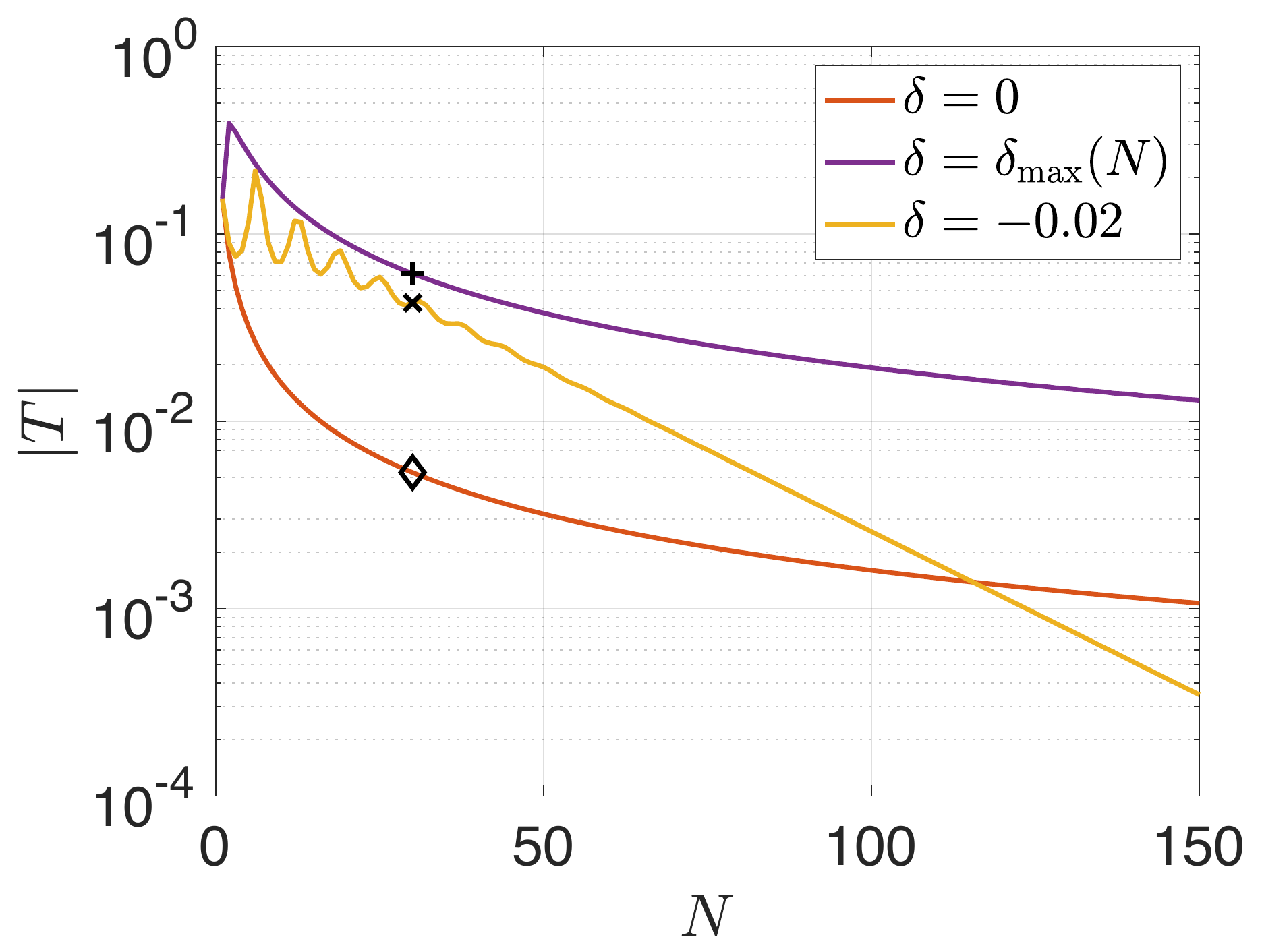}
\caption{Transmission over a $N$ cells of the configuration of Fig.~\ref{ResAndMesh_series_Fig} with resistive sheets in the large cross-section tube as a function of the frequency centered on the lower edge of the gap $\delta = k - k_\star$. We used $\gam=2$, $\nu = 40$, $\ell_1/\ell=0.5$. $\delta_\M(N)$ is the frequency shift giving the maximum transmission.}
\label{Scatt_nearPeak_WMinPeriodic_Nscale_Fig} 
\end{figure}

To better understand the properties of the transmission peak, and in particular its maximum value, we analyze the transmission in the vicinity of the EP. We show this in Fig.~\ref{Scatt_nearPeak_WMinPeriodic_Nscale_Fig} for the configuration where the anomalous transmission is at the lower edge of the first gap, at a frequency noted $k=k_\star$. The first thing to notice is that the peak is asymmetric: frequencies above the EP are highly suppressed because they correspond to the location of the gap of the base medium. On the contrary, frequencies below the EP are less efficiently suppressed. Moreover, the transmission oscillates while decreasing when the frequency moves away below the EP. As a result, the maximum transmission is not reached at the EP, but at a frequency close below, and the transmission value is always higher than that at the EP (see Fig.~\ref{Scatt_nearPeak_WMinPeriodic_Nscale_Fig} left panel). When one looks at the scaling of the transmission as a function of $N$, the behavior is similar to the array of resistive sheets: at a fixed frequency, the transmission decreases exponentially with $N$, except at the EP, where it decreases linearly. However, the maximum transmission is reached at a frequency $k-k_\star = O(1/N)$, which gets closer to the EP when $N$ increases. This lead to a maximum transmission also decreasing linearly as a function of $N$ (see Fig.~\ref{Scatt_nearPeak_WMinPeriodic_Nscale_Fig} right panel). 

As a last remark, we point out that the symmetry of the wave function at the edges of a gap is directly related to the topological properties of the system~\cite{Ma19}. In particular, a noticeable change occurs when the base medium undergoes a topological phase transition. Indeed, it was observed in~\cite{Novikov19} that the location of the anomalous transmission peaks changes across a topological phase transition. Such a transition is characterized by a gap closing and reopening when varying a parameter (such as the cross-section ratio $\nu$ here), while the wave functions at the edges of the gap exchange symmetry~\cite{Xiao14}. Therefore, from what precedes, we conclude that the anomalous transmission frequency will change from the lower edge of the gap to the upper edge of the gap, in agreement with what was observed in~\cite{Novikov19}.

\section{Conclusion}

In this work, we use a periodic set of resistive sheets to realize an acoustic analogue of the Borrmann effect. This effect manifests itself as an anomalously high transmission at the Bragg frequency of the system. This anomalous transmission is theoretically predicted for our system and observed experimentally. Using the transfer matrix formalism, we show that this anomalous transmission corresponds to an exceptional point at the Bragg frequency. As a result, the transmission coefficient decrease linearly with the number $N$ of sheets, while it decreases exponentially for other frequencies. This change of scaling implies that at large $N$, the transmission shows a very narrow peak at the Bragg frequency. In addition, we show that the subwavelength size of the resistive sheets is a key ingredient. When taking into account a finite thickness, the exceptional point is replaced by an avoided crossing, and the maximum transmission is lowered accordingly. 

Moreover, the exceptional point approach of the effect allows us to show that a similar transmission peak can be obtained at the edge of a band-gap of a lossless periodic medium by placing resistive sheets at the velocity nodes of the eigensolution of the problem. This complements similar observations in photonic crystals~\cite{Vinogradov09,Novikov17,Novikov19} by providing a constructive way of obtaining a Borrmann anomalous transmission and relating it to the topological properties of the system.

\newpage
\appendix
\section{Transfer matrices and scattering coefficients}
\label{Mmat_Smat_App}
\begin{figure}[!ht]
\centering
\includegraphics[width=0.49\columnwidth]{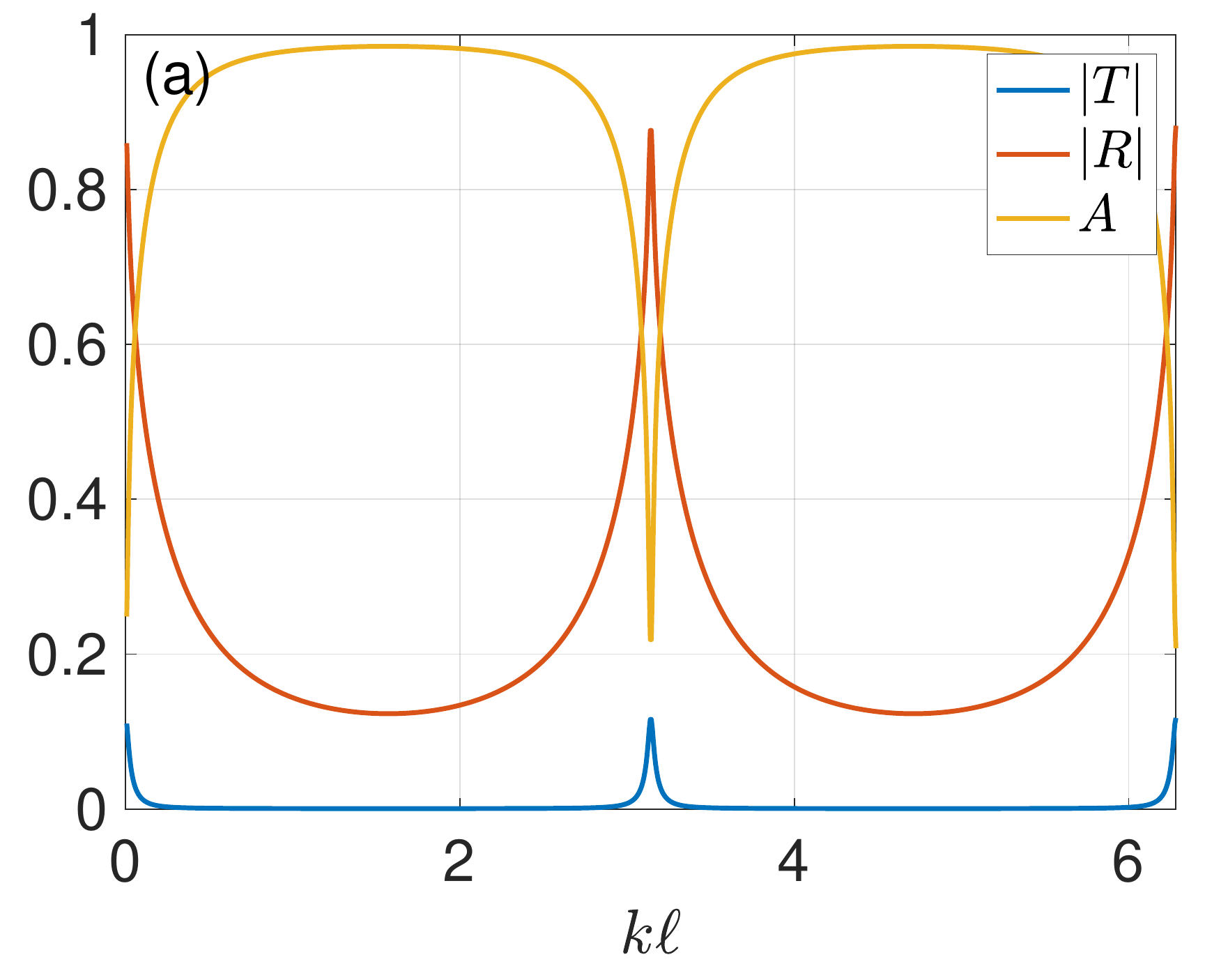}
\includegraphics[width=0.49\columnwidth]{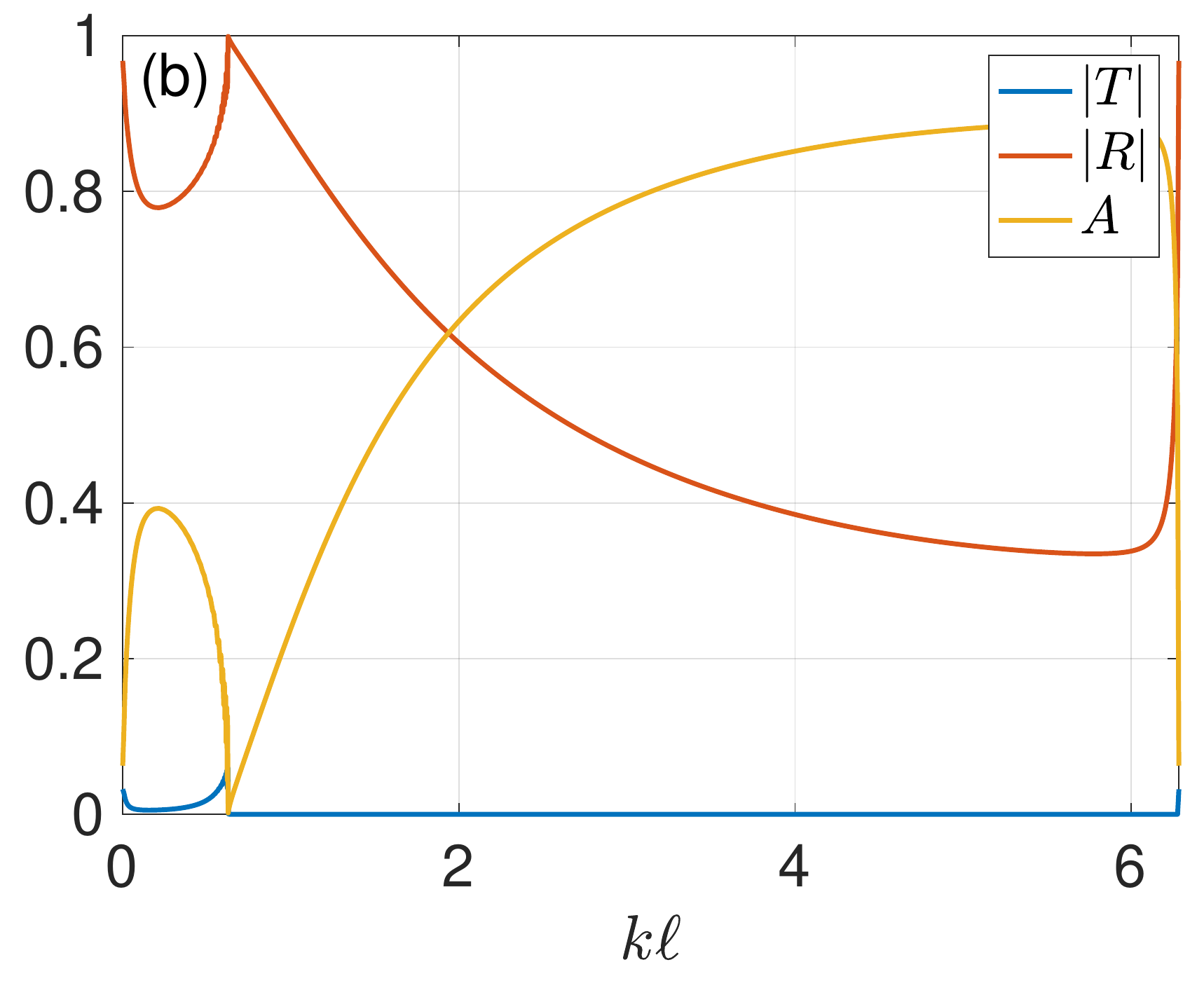}
\caption{(a) scattering coefficients over $N$ equidistant resistive sheets as in Fig.~\ref{WM_Schema_Fig}. We used $N=30$ and $\gam=0.5$. (b) scattering coefficients over $N$ cells of resonators and resistive sheets in the middle of the large cross section tube (configuration of Fig.~\ref{ResAndMesh_series_Fig}~(a)). We used $N=30$, $\nu=40$, $\ell_1=\ell_0$, and $\gam = 2$.}
\label{Scatt_coefs_Fig} 
\end{figure}

In this section we give explicit formulas for the scattering coefficients over a slab of $N$ cells. For this we use the Chebyshev identity~\cite{Soukoulis}, which gives the $N$-th power of a $2 \times 2$ matrix. We first define the transfer matrix by writing the pressure field as a sum of forward and backward waves $p(x) = p_+(x) + p_-(x)$~\cite{Soukoulis}. Each component are easily expressed in terms of pressure and velocity:
\be
p_+ = \frac{p+u}{2}, \qquad \textrm{and} \qquad p_- = \frac{p-u}{2}, 
\ee
so that in the absence of scatterer $p_\pm(x) = A_\pm e^{\pm i k x}$. Then, the transfer matrix from a point $x_1$ to a point $x_2$ is defined by 
\be
\bmat p_+(x_2) \\ p_-(x_2) \emat = M \cdot \bmat p_+(x_1) \\ p_-(x_1) \emat. 
\ee
We now consider the general form of a transfer matrix for the unit cell:
\be \label{General_Mmat}
M_c \doteq \bmat \alpha & \beta \\ \tilde \beta & \tilde \alpha \emat, 
\ee
We assume reciprocity (which gives $\det(M) = 1$), and hence, we can write the eigenvalues of $M_c$ as $\lam = e^{\pm i q \ell}$. The $N$-th power of $M_c$ is given by 
\be \label{Cheb_ident}
M_c^N = \bmat \frac{\sin(N q \ell)}{\sin(q \ell)} \alpha - \frac{\sin((N-1)q \ell)}{\sin(q \ell)} & \frac{\sin(N q \ell)}{\sin(q \ell)} \beta \\
\frac{\sin(N q \ell)}{\sin(q \ell)} \tilde \beta & \frac{\sin(N q \ell)}{\sin(q \ell)} \tilde \alpha - \frac{\sin((N-1)q \ell)}{\sin(q \ell)} \emat . 
\ee
We now consider the scattering on an array of resistive sheets made of $N$ unit cells, see Fig.~\ref{WM_Schema_Fig}. Using this and the relation between the transfer matrix and the scattering coefficients \eqref{tot_Smat}, we obtain 
\bsub \label{Borr_General_Scatt} \bea
\frac1T &=& \left(1 + \dfrac{\gam}{2}\right) \dfrac{\sin(N q \ell)}{\sin(q \ell)} e^{-i k \ell} - \dfrac{\sin((N-1) q \ell)}{\sin(q \ell)} , \\
R &=& -\dfrac{\gam}{2} \dfrac{\sin(N q \ell)}{\sin(q \ell)} e^{-i k \ell} \times T .
\eea \esub
Evaluating this at the Bragg frequency $k\ell = q\ell = \pi$ gives us the maximum transmission \eqref{T_max}. To evaluate the plateau of minimum transmission, we evaluate it at the frequency $k\ell = \pi/2$, as in the core of the text. First, we see from the dispersion relation \eqref{Borr_DispRel} that the Bloch wavenumber is given by 
\be
q\ell = \frac\pi2 + i \mathrm{argsh}\left(\frac{\gam}2\right). 
\ee
Then, using it in the above expressions \eqref{Borr_General_Scatt} we obtain 
\be
\frac1{|T_\m|} = \left( \frac{1+\sqrt{1+\gam^2/4}}{2\sqrt{1+\gam^2/4}} \right)e^{N \mathrm{argsh}\left(\gam/2\right)} + (-1)^{N+1} \left( \frac{1-\sqrt{1+\gam^2/4}}{2\sqrt{1+\gam^2/4}} \right)e^{-N \mathrm{argsh}\left(\gam/2\right)}. 
\ee
Taking the limit $N \gg 1$, we obtain \eq{T_min}. We also give the transfer matrix of a cell of the array of resonators (Fig.~\ref{ResAndMesh_series_Fig}) in the monomode approximation: 
\be
M_c^{(0)} = \bmat \frac{(1+\nu)^2}{4\nu} e^{i k (\ell_1+\ell_0)} - \frac{(1-\nu)^2}{4\nu} e^{i k (\ell_0 - \ell_1)} &
i\frac{1-\nu^2}{2\nu} \sin(k \ell_1) \\ 
-i\frac{1-\nu^2}{2\nu} \sin(k \ell_1) & 
\frac{(1+\nu)^2}{4\nu} e^{-i k (\ell_1+\ell_0)} - \frac{(1-\nu)^2}{4\nu} e^{-i k (\ell_0 - \ell_1)} \emat . 
\ee
The superscript $(0)$ is here to recall that it is lossless. When $\nu>1$, $M_c^{(0)}$ describes Fig.~\ref{ResAndMesh_series_Fig}~(a), while when $\nu < 1$ it describes Fig.~\ref{ResAndMesh_series_Fig}~(b). To add the resistive sheets, we simply multiply $M_c^{(0)}$ by the transfer matrix of a single resistive sheet $M_\gam$ on the right. The latter is given by \eq{Mmat_cell} with $\ell=0$: 
\be
M_\gam = \bmat \vspace{8pt} 1 - \dfrac{\gam}{2} & \dfrac{\gam}{2} \\ - \dfrac{\gam}{2} &  1 + \dfrac{\gam}{2} \emat. 
\ee
In the core of the text, we mainly focused on transmission properties of the different media. But it is also instructive to look at the reflection coefficient and the absorption $A = 1 - |T|^2 - |R|^2$. This is done in Fig.~\ref{Scatt_coefs_Fig} for the array of resistive sheet (configuration of Fig.~\ref{WM_Schema_Fig}) and the resistive sheets inside the array of resonators (configuration Fig.~\ref{ResAndMesh_series_Fig}). Notice that in both cases, the increase of transmission due to the Borrmann effect is also accompanied by an increase in reflection, and therefore, a decrease of absorption.

\section{General condition to obtain an exceptional point with resistive sheets}
\label{General_EP_App}
In Sec~\ref{MeshandRes_Sec}, we saw how one can obtain an anomalous transmission using a base medium (lossless) and adding resistive sheets. Although we used in this work a base medium made of series resonators, the construction of EPs in terms of the symmetry of the eigenvector of the cell transfer matrix is very general. To illustrate this, we consider a base medium described by a transfer matrix of the form \eqref{General_Mmat}. We only assume reciprocity and mirror symmetry, which impose 
\bea 
\beta + \tilde \beta &=& 0 , \label{EP_cond1} \\
\alpha \tilde \alpha + \beta^2 &=& 1. \label{EP_cond2}
\eea
Now, since $2\cos(q\ell) = \mathrm{Tr}(M)$, on the edge of the first gap (the construction is similar for higher gaps) we must have $\mathrm{Tr}(M) = -2$. This gives us a last relation
\be \label{EP_cond3}
\alpha + \tilde \alpha = -2. 
\ee
We can now solve equations \eqref{EP_cond1}, \eqref{EP_cond2} and \eqref{EP_cond3}, and we obtain the general form of the transfer matrix at an exceptional point:  
\be \label{BG_edge_Mmat}
M_c^{(0)} = \bmat \vspace{8pt}  -1 - i\dfrac{\mu_0}2 & i s_\pm \dfrac{\mu_0}2 \\ - i s_\pm \dfrac{\mu_0}2 & -1 + i \dfrac{\mu_0}2 \emat, 
\ee
where $\mu_0$ is defined by $d-a= i \mu_0$, and $s_\pm = \pm 1$. Notice that because the base medium is assumed lossless, $\mu_0 \in \mathbb R$. The important factor is $s_\pm$, because this sign dictates the symmetry of the eigenvector. Indeed, we see that the (unique) eigenvector of the matrix \eqref{BG_edge_Mmat} is 
\be
\bmat A \\ B \emat = \bmat 1 \\ s_\pm \emat . 
\ee
If $s_\pm = 1$, the eigenvector is symmetric, if $s_\pm = -1$, the eigenvector is anti-symmetric. Since the transfer matrix $M_\gam$ of a single resistive sheet is at an EP, it commutes with $M_c^{(0)}$ if and only if they have the same eigenvector. Since the eigenvector of $M_\gam$ is symmetric, we see that if $s_\pm = 1$, the product $M_c^{(0)} \cdot M_\gam$ is at an EP. Translated in terms of wave solutions, this symmetry condition means that the resistive sheets must be placed on the velocity nodes of the eigen-solution of $M_c^{(0)}$.

\bibliographystyle{utphys}
\bibliography{Bibli}

\end{document}